\documentclass[%
 preprint,aps,
amsmath,amssymb,
superscriptaddress,
aps, 11pt]{revtex4-2}

\usepackage{amsmath}
\usepackage{amssymb}
\usepackage{dcolumn}
\usepackage{bm}
\usepackage{hyperref}
\usepackage{verbatim}
\usepackage{color}
\usepackage{xcolor}
\usepackage[version=4]{mhchem}
\usepackage{verbatim}
\usepackage{graphics}
\usepackage{lipsum}
\usepackage{graphicx}

\usepackage{subfigure}
\usepackage{subcaption}
\usepackage{wrapfig}
\usepackage{epsfig}
\usepackage{float}
\usepackage{array}
\usepackage{psfrag}
\usepackage{color}
\usepackage{bbold}
\newcommand{\vect}[1]{\boldsymbol{#1}}

\definecolor{azure}{rgb}{0.0, 0.5, 1.0}
\definecolor{asparagus}{rgb}{0.53, 0.66, 0.42}
\definecolor{ballblue}{rgb}{0.13, 0.67, 0.8}
\definecolor{sgreen}{rgb}{0.0, 0.8, 0.35}
\definecolor{darkgreen}{rgb}{0.0, 0.5, 0.0}
\definecolor{sred}{rgb}{0.9, 0.6, 0.4}
\usepackage{textcomp}

\usepackage{tikz}

\usepackage[]{color}

\begin{document}

\thispagestyle{plain}



\makeatletter 
\renewcommand\@biblabel[1]{#1}            
\renewcommand\@makefntext[1]%
{\noindent\makebox[0pt][r]{\@thefnmark\,}#1}
\makeatother 
\renewcommand{\figurename}{\small{Fig.}~}
\setlength{\skip\footins}{0.8cm}
\setlength{\footnotesep}{0.25cm}
\setlength{\jot}{10pt}

\setlength{\arrayrulewidth}{1pt}
\setlength{\columnsep}{6.5mm}
\setlength\bibsep{1pt}

\makeatletter 
\newlength{\figrulesep} 
\setlength{\figrulesep}{0.5\textfloatsep} 

\makeatother

\title{
Theory of Wetting Dynamics with Surface Binding
}

\author{{Xueping Zhao}}
\affiliation{Department of Mathematical Sciences, University of Nottingham Ningbo China, Taikang East Road 199, 315100 Ningbo, China}

\author{Susanne Liese}
\affiliation{
 Faculty of Mathematics, Natural Sciences, and Materials Engineering: Institute of Physics, University of Augsburg, Universit\"atsstra\ss e~1, 86159 Augsburg, Germany
}

\author{Alf Honigmann}
\affiliation{Biotechnology Centre (BIOTEC), TU-Dresden, Tatzberg 47, 01307 Dresden, Germany}
\affiliation{
Cluster of Excellence Physics of Life, TU Dresden, 01062 Dresden, Germany 
}

\author{Frank Jülicher}
\affiliation{
Max Planck Institute for the Physics of Complex Systems, Nöthnitzer Stra\ss e~38, 01187 Dresden, Germany
}
\affiliation{
Center for Systems Biology Dresden,  Pfotenhauerstra\ss e~108, 01307 Dresden, Germany}
\affiliation{
Cluster of Excellence Physics of Life, TU Dresden, 01062 Dresden, Germany 
}
\affiliation{
corresponding authors: julicher@pks.mpg.de and christoph.weber@physik.uni-augsburg.de
}

\author{Christoph A. Weber}
\affiliation{
 Faculty of Mathematics, Natural Sciences, and Materials Engineering: Institute of Physics, University of Augsburg, Universit\"atsstra\ss e~1, 86159 Augsburg, Germany
}
\affiliation{
corresponding authors: julicher@pks.mpg.de and christoph.weber@physik.uni-augsburg.de
}


\begin{abstract}

Biomolecules, such as proteins and RNAs, 
can phase separate in the cytoplasm of cells to form biomolecular condensates. Such condensates are liquid-like droplets that can wet biological surfaces such as membranes.
Many molecules that participate in phase separation can also reversibly bind to membrane surfaces.
When a droplet wets a surface, molecules can diffuse inside and outside of the droplet or in the bound state on the surface.
How the interplay between surface binding, diffusion in surface and bulk affects the wetting kinetics is not well understood. 
Here, we derive the governing equations using non-equilibrium thermodynamics by relating the thermodynamic fluxes and forces at the surface coupled to the bulk. 
We study the spreading dynamics in the presence of surface binding and find that binding speeds up wetting by nucleating a droplet inside the surface. 
Our results suggest that the wetting dynamics of droplets can be regulated by two-dimensional surface droplets in the surface-bound layer through changing the binding affinity to the surfaces. These findings are relevant both to engineering life-like systems with condensates and vesicles, and biomolecular condensates in living cells.

\end{abstract}

\maketitle

\newpage



\section{Introduction}

Living cells organize their chemical reactions in space by forming various compartments. 
These compartments provide different chemical environments for distinct biochemical processes. 
Some compartments are bounded by a membrane surface composed of lipids and proteins. 
Examples are the nucleus~\cite{Boisvert2007}, endoplasmic reticulum, 
golgi apparatus, 
mitochondria~\cite{Friedman_Nunnari_Nature_2014}, plastids, 
lysosomes~\cite{Luzio_2007} and
endosomes. 
However, many compartments in cells have no membrane as boundaries. 
Examples include the nucleolus~\cite{Brangwynne_Hymann_PNAS_2011},  centrosomes~\cite{Mahen&Venkitaraman2012}, 
Cajal bodies~\cite{Gall2003},
P granules~\cite{Brangwynne2009, fritsch2021local}, and stress granules~\cite{Buchan&Parker2009, Decker&Parker2012}.  These membrane-less organelles, termed biomolecular condensates,  are liquid-like droplets formed in a process similar to liquid-liquid phase separation~\cite{Brangwynne2009,Brangwynne_Hymann_PNAS_2011,fritsch2021local}. 

Biomolecular condensates can attach to  biological surfaces such as membranes. 
This process is referred to as wetting. Examples are P granules wetting on the surface of the nucleus~\cite{Brangwynne2009}, or TIS granules wetting the
endoplasmic reticulum~\cite{Ma_sayr_Cell_2018}.
Furthermore, many biological molecules that  phase separate can also bind to biological surfaces, leading to a two-dimensional molecular layer on the surface. The molecules in such surface layers can give rise to rich spatio-temporal patterns. For example, the PAR proteins bind and unbind to the surface periodically, leading to asymmetrical localization during the asymmetric cell division~\cite{Goehring2011, Lars2019}. The \textit{Escherichia-coli} MinCDE system is another example of pattern formation in the surface layer of bound molecule layer~\cite{Ramm_Schwille_2019}.
Molecules bound to surfaces can also nucleate the formation of biomolecular condensates in the bulk. An example is Sec bodies induced by amino-acid starvation in Drosophila cells where  small condensate nuclei form at the endoplasmic reticulum's exit sites (ERES)~\cite{Zacharogianni_elife_2014}.

The relevance of surface binding and clustering for surface phase transitions such as surface phase separation and prewetting was studied at thermodynamic equilibrium~\cite{NJPpaper2021, rouches2022surface, julicher2023droplet}.
A key finding is that the 
prewetting transition can occur far below the equilibrium concentration and is accessible for a larger range of thermodynamic parameters. 
Ref.~\cite{NJPpaper2021} also showed that the interplay between surface-bound molecules and free diffusive molecules in the bulk can also lead to multiple pre-wetted states.
Interestingly, binding also shifts the wetting transition line and affects the contact angle.   
{These phenomena rely on surface binding, effectively modifying the properties of the surface for prewetting and wetting. 
The idea of modifying surface properties to affect  wetting  was also explored in systems undergoing reactive~\cite{Kumar_2007,Eustathopoulos2006} or adaptive wetting~\cite{Butt2018}.}

The droplet dynamics of wetting 
were described using a fluid dynamical model based on viscous dissipation and a transition state model describing the contact line motion as adsorption and desorption kinetics~\cite{Gennes1984, gennes2004capillarity}. 
The dynamics toward a completely wetted state are governed by  Tanner's law~\cite{gennes2004capillarity, cormier2012beyond}, 
where the contact angle exhibits a power-law relaxation in time $t$ toward thermodynamic equilibrium, $\theta\propto t^{-3/10}$.
This law implies very slow spreading dynamics of the droplet contact area proportional to $t^{1/5}$. 
Tanner's law and the dynamics of the contact area require a $nm$-thick precursor film
on which the  droplet spreads~\cite{hardy1919boundary, leger1988precursor, xu2004molecular, popescu2012precursor}.
The film extends from the droplet further out and grows with the wetting droplet. 
In biological systems, such a film of molecular thickness could form
via the binding of molecules from the liquid bulk domains to the adjacent biological surfaces. 
Despite the significance of wetting and molecular binding in biological systems~\cite{julicher2023droplet,bulk_surface_exp}, 
the exploration of the dynamics of wetting in conjunction with surface binding, remains largely unexplored. 

To bridge this gap, we have derived the governing dynamic equations of a bulk droplet wetting a surface where  droplet molecules can bind. To this end, we use irreversible thermodynamics. Moreover, we developed a two-dimensional numerical solver to explore the effects of surface binding on the dynamics of wetting. We show that  surface binding speeds up the dynamics of droplet spreading by orders in magnitude depending on the binding rate coefficient. 
Our findings indicate that surface binding can act as a switch to control the wetting of droplets in living cells.

This paper is structured as follows: In Section~\ref{sect:theory}, we derive the model for phase separation in bulk and surface coupled by surface binding using irreversible thermodynamics. 
Section~\ref{sect:wettingwithsurfacebinding} is devoted to an application of the developed dynamic model: We discuss spreading to a completely wetted state and spreading toward a partially wetted state. 
Finally, in Section~\ref{sect:conclusion_outlook}, we present our conclusions and outline potential future directions based on our findings.

\begin{figure*}
\centering
{\includegraphics[width=1\textwidth]{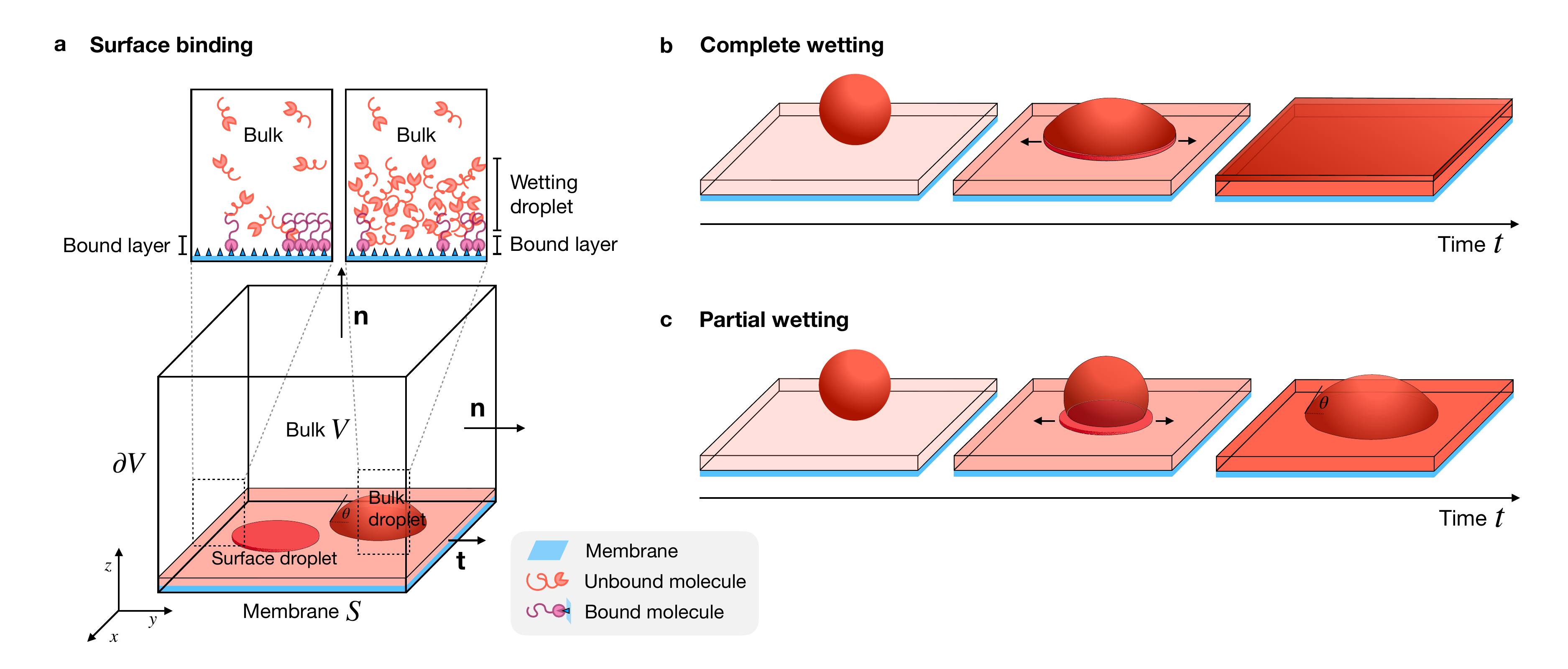}}
\caption{\textbf{Schematics of wetting dynamics on binding surfaces.
}
\textbf{a)} 
A binary mixture in bulk and surface is shown to be coupled by interactions and surface binding. The bulk mixture can phase separate and give rise to a bulk droplet. Binding builds up a molecular layer of bound molecules that can phase-separate, leading to a surface droplet. We consider a planar binding surface $S$ and a cubic bulk domain $V$. 
The non-interacting and non-binding surfaces are denoted as $\partial V$. 
Wetting of the bulk droplet on the surface is characterized by the contact angle $\theta$.
\textbf{b,c)} We study the spreading dynamics of a bulk droplet towards a completely wetted or partially wetted thermodynamic equilibrium state. 
}
\label{fig:sketch}
\end{figure*}

\section{Non-equilibrium thermodynamics 
of wetting with surface binding}
\label{sect:theory}

We consider a binary mixture in a three-dimensional domain $V$ that we refer to as the bulk in the following. 
The bulk is coupled to a two-dimensional surface $S$.
We describe the system using a canonical ensemble where  temperature $T$, 
the size of the volume $|V|$ and surface area $|S|$,  as well as
the total particle number $N$ in bulk and surface are fixed. 
For simplicity, the bulk is a cubic volume $|V|=L^3$, where $L$ is the side length. The molecules in the bulk can bind to the lower surface $S$ of the cubic volume, which we refer to as the binding surface. The remaining non-binding and non-interacting surfaces of the cubic volume are denoted by $\partial V$; see Fig.~\ref{fig:sketch}a for a sketch. 
The molecules in the bulk domain $V$ diffuse, bind to, and unbind from the surface $S$, leading to a layer of bound molecules. Molecules in this layer  can only diffuse on the surface. 
We consider an incompressible system in bulk and surface, respectively. 
Thus, we can describe the wetting dynamics depicted in Fig.~\ref{fig:sketch}b,c using bulk volume fraction $\phi(\vect{x},t)$ of the molecules and their area fraction bound to the surface, $\phi_s(\vect{x}_\parallel,t)$.  Here, $\vect{x}=(x,y,z)$ is the bulk position and $\vect{x}_\parallel=(x,y)$ the position in the surface.
The surface is located at $z=0$.

\subsection{Conservation laws}

The dynamics of the bulk volume fraction $\phi$ and the area fraction $\phi_s$  are governed by conservation laws for surface and bulk: \begin{subequations}
\label{eq:cons_laws_full_sodel}
\begin{align}
\partial_t \phi_{} &= -\nabla \cdot \vect{j}_{}\, , & \vect{x} &\in V \, ,
\\
\partial_t \phi_s &= -\nabla_{\parallel} \cdot \vect{j}_s + r \, , &  \vect{x} & \in S \, , 
\end{align}
where $\vect{j}$ and $\vect{j}_s$ are the diffusive fluxes in bulk and surface. Moreover, $r$ denotes the binding flux. 
Particle conservation relates the binding flux $r$ to the normal component of the diffusive flux $\vect{j}$ at the surface (derivation see Appendix~\ref{APP:binding_flux}):
\begin{align}\label{eq:bc_flux_binding}
	 \vect{n} \cdot  \vect{j} = r \,  \frac{\nu}{\nu_s} \, , \quad \vect{x}  \in S \, ,
\end{align}
where $\nu$ and $\nu_s$ denote the molecular volume and molecular area, and $ \vect{n}$ denotes the surface normal pointing outward of the volume domain $V$; see Fig.~\ref{fig:sketch}. 
As molecules cannot bind or interact with the remaining boundaries $\partial V$, the flux normal to the surface vanishes:
\begin{align}\label{eq:bc_flux_nobinding}
	 \vect{n} \cdot  \vect{j} = 0 \, , \quad \vect{x}  \in \partial V \, .
\end{align}
\end{subequations}
Alternatively, for the $\partial V$ boundaries that are not opposite to the binding surface $S$, periodic boundaries could be considered. 
We also have to impose the conditions at the one-dimensional boundaries of the binding surface $S$, which we denote as $\partial S$. To conserve volume, periodic boundary conditions could be imposed or
$\vect{t} \cdot \vect{j}_s=0$, where $\vect{t}$ is the normal to the one-dimensional boundary $\partial S$; see Fig.~\ref{fig:sketch}.
The dynamic equations and boundary conditions above conserve the total particle number $N$ in bulk and in the surface, 
\begin{equation}
\label{eq:tot_N}
    N= \frac{|V|}{\nu} \bar{\phi}  + \frac{|S|}{\nu_s} \bar{\phi_s}  \, , 
\end{equation}
where 
\begin{equation}
    \bar{\phi}(t) = |V|^{-1}\int_V dV \phi(\vect{x},t)
\end{equation}
is the average bulk volume fraction and
\begin{equation}
\label{eq:average_area_fraction}
\bar{\phi}_s(t) = |S|^{-1} \int_S dS \, \phi_s(\vect{x}_\parallel,t)
\end{equation}
is the average area fraction. 
Moreover,
$|V|$ is the volume of the bulk, and $|S|$ is the surface area of the surface to which the molecules bind.

The fluxes $\vect{j}$, $\vect{j}_s$, and $r$ are driven by the conjugate thermodynamic forces. These relationships are derived in Section~\ref{sec:IrrevThermod1} using irreversible thermodynamics. Since the thermodynamic forces are non-linear in the field $\phi(x,t)$ and $\phi_s(x,t)$, we employ numerical methods to solve the non-linear dynamic equations. 

\subsection{Free energy}

{The free energy governs interactions among components in bulk and surface. The total Helmholtz free energy $F[\phi,  \phi_s]$ depends on the three independent thermodynamic fields 
$\phi(\vect{x},t)$ and $\phi_s(\vect{x}_\parallel,t)$ and their spatial derivatives.
It can be decomposed of four parts: the bulk free energy density $f(\phi)$, the surface free energy density $f_s(\phi_s)$, the coupling free energy density between bulk and surface $J(\phi|_{z=0}, \phi_s)$, and free energy costs for gradients of the bulk volume fraction parallel to the surface:}
\begin{align}\label{eq:free_energy_F}
 F[\phi,  \phi_s]  &= \int_{V} dV \Big [ f(\phi) + \frac{1}{2}\kappa \left(\nabla \phi\right)^2 \Big ]  \\ \nonumber 
    & \quad  +  \int_S dS \Big [ f_s(\phi_s) + \frac{1}{2}\kappa_s \left(\nabla_{\parallel} \phi_s\right)^2 + J(\phi_s, {\phi|_{z=0}} )  +  \frac{1}{2}\kappa_{0}\left(\nabla_{\parallel} {\phi|_{z=0}}\right)^2 \Big ] \, ,
\end{align}
where $\phi|_{z=0}(\vect{x}_\parallel,t) = \phi(x,y,z, t)|_{z=0}$ is the bulk volume fraction at the surface, and $dV=dxdydz$ and $dS=dxdy$ are the volume and surface elements for the considered cubic system.  
The free energy costs due to gradients in volume fraction in bulk and at the surface are characterized by the coefficients $\kappa$, $\kappa_0$, respectively, while in the bound layer, the correspondingly parameter is $\kappa_s$. Moreover, $\nabla_\parallel=(\partial_x,\partial_y)$ denotes the gradient vector in the $x$-$y$ surface plane. 

Before we derive the diffusive fluxes $\vect{j}$ and $\vect{j}_s$ and the binding flux $r$ using irreversible thermodynamics in Sect.~\eqref{sec:IrrevThermod1}, we determine the conditions for thermodynamic equilibrium as a reference in the next section. 
 
\subsection{Thermodynamic equilibrium }\label{sec:Thermod1}

At thermodynamic equilibrium and for a $T$-$V$-$N$ ensemble, the total Helmholtz free energy $F$ is minimal
with the binding constraint of the total molecule number $N$ (Eq.~\eqref{eq:tot_N}) being conserved:
\begin{equation}
\label{eq:thermodyn_eq_condition}
0=\delta \left( F[\phi,  \phi_s]- 
\lambda \left[ \int_V   dV  \,  \phi_{}  /\nu  +  \int_S dS\, \phi_s  /\nu_s    - N_{} \right] \right) \, ,
\end{equation}
where $\lambda$ is the Lagrange multiplier fixing  $N$.
Using Eq.~\eqref{eq:free_energy_F}, the variation of the total Helmholtz free energy is given as:
\begin{align}
\nonumber
\delta F &= 
\int_V dV \, \Big(    f_{}^\prime \left({\phi_{}}\right)  -  \kappa_{}  \nabla^2  \phi_{}   \Big) \delta \phi_{} 
 \\
&\quad  
+ \int_S dS  \, \left[ \Big (   f_s^\prime \left({\phi_s}\right)  -  \kappa_s \nabla_{\parallel}^2 \phi_s  +  \frac{\partial J}{\partial \phi_s}  \Big) \delta \phi_s  + 
\, \Big( \frac{\partial J}{\partial {\phi|_{z=0}}}  - \kappa_0 \nabla_{\parallel}^2 {\phi|_{z=0}} +   \vect{n} \cdot \kappa_{}\left(\nabla \phi_{} \right){|_{z=0}} \Big){\delta \phi|_{z=0}}
\right] 
\nonumber\\
&\quad +\int_{{\partial V}} dS \, 
\Big(  \vect{n} \cdot \kappa_{} \left( \nabla \phi\right) \Big)  {\delta \phi|_{\partial V}}
+ \int_{\partial S} d l \,  \Big (   \vect{t} \cdot \kappa_s \nabla_{\parallel} \phi_s  \Big ) \delta \phi_s    
\, ,
\label{eq:variation_F}
\end{align} 
where  $\partial S$ is the one-dimensional boundary of the surface $S$ and $\vect{t}$ is the {normal} vector to this boundary.  

{We identify five thermodynamic forces related to deviations of the independent fields $\phi$ and $\phi_s$ in the respective spatial domains. 
Following Eq.~\eqref{eq:variation_F}, 
we define the exchange chemical potentials for the bulk, $\mu$, 
exchange chemical potentials for the binding to the surface, $\mu_s$,
and the chemical potential for the surface of the bulk boundary, $\mu_0$:}
\begin{subequations}\label{eq:chem_pot_def}
\begin{align}
\label{eq:EL_bulk}
\mu_{} &={ \nu}  \frac{\delta F}{\delta \phi_{}} =\nu \Big( f_{}^\prime \left( \phi_{} \right) - \kappa_{}  \nabla^2  \phi_{}\Big)
\\
\label{eq:chem_pot_surface}
\mu_s &={ \nu_s}  \frac{\delta F}{\delta \phi_s} =  \nu_s \left(  f_s^\prime \left( \phi_s \right) -  \kappa_s \nabla_{\parallel}^2 \phi_s  + \frac{\partial J}{\partial \phi_s}  \right) \, ,
\\
\mu_{0} &= \nu_s \frac{\delta F}{\delta \phi|_{z=0}} = \nu_s \Big( \frac{\partial J}{\partial {\phi|_{z=0}}}  - \kappa_0 \nabla_{\parallel}^2 {\phi|_{z=0}} +    \vect{n} \cdot \kappa_{} \left(\nabla \phi_{} \right){|_{z=0}} \Big)
\label{eq:chem_pot_bulk_surface}
\, ,
\end{align}
\end{subequations}
where the prime denotes a  derivative, e.g., $f_{}^\prime (\phi) = d f/d\phi$.
{The bulk chemical potential $\mu$ is related to variations of the volume fraction in the bulk, $\delta \phi$, 
the surface chemical potential $\mu_s$ to variations of the area fraction of bound molecules, $\delta \phi_s$,
and $\mu_{0}$ to variations of the bulk volume fraction at $z=0$ at the binding surface $S$, $\delta \phi|_{z=0}$. 
Similarly, $\vect{n} \cdot \kappa_{} \left( \nabla \phi\right){|_{z=0}}$ is the thermodynamic force associated with  deviations of  bulk volume fraction at the non-interacting and non-binding surface $\partial V$, $\delta \phi|_{\partial V}$. 
Finally, $  \vect{t} \cdot  \kappa_s \nabla_{\parallel} \phi_s$ is a thermodynamic force when perturbing $\phi_s$ at the boundary of $S$, denoted as $\partial S$.
We note that the bulk chemical potential evaluated at $z=0$
is different from the chemical potential for the surface of the bulk boundary $\mu_0$ ($\mu|_{z=0}\not=\mu_0$). 
}

{All five thermodynamic forces 
characterize the work performed when varying one of the concentration fields in a specific spatial domain. 
When changing this field away from its equilibrium value, dissipation occurs and entropy is produced, which we discuss in detail in the next section on irreversible thermodynamics~\ref{sec:IrrevThermod1}.}

The entire system composed of surface and bulk is at thermodynamic equilibrium if  condition~\eqref{eq:thermodyn_eq_condition} is satisfied. This condition implies that the surface and bulk exchange chemical potentials are constant and equal to the Lagrange multiplier,
\begin{subequations}\label{eq:equilibrium}
\begin{align}
	\lambda &= \mu_s=\mu_{} \, .
\end{align}
Moreover, the following boundary conditions 
at the surface $S$ and the other non-binding and non-interacting surfaces $\partial V$ of the volume $V$ fulfill:
\begin{align}
\label{eq:bc_EQa}
\kappa \, \vect{n} \cdot  \left( \nabla  \phi \right){|_{z=0}} &= -  \frac{\partial J}{\partial{ \phi|_{z=0}}} + \kappa_0 \nabla_{\parallel}^2 \phi{|_{z=0}} \, , &  \vect{x} & \in S \, , \\
\label{eq:bc_EQb}
\kappa \, \vect{n} \cdot \left( \nabla  \phi \right){|_{\partial V}} &= 0  \, , &  \vect{x} & \in \partial V  \, ,\\
\label{eq:bc_EQc}
\kappa_s \, \vect{t} \cdot   \nabla_{\parallel} \phi_s &= 0  \, , &  \vect{x} & \in \partial S  \,  
. 
\end{align}
\end{subequations}
Note the $\partial V$ boundaries are termed `non-interacting' since the coupling free energy density $J$ vanishes at such boundaries. 
They are also called `non-binding' because either the flux normal to $\partial V$ vanishes (Eq.~\eqref{eq:bc_flux_nobinding}) or periodic conditions are considered at $\partial V$. 

The boundary conditions Eqs.~\eqref{eq:bc_EQa}-\eqref{eq:bc_EQc} are related to the contact angle $\theta$ which is defined in the limit of a sharp interface~\cite{liese2023chemically}. In this limit, the width of the interface profile $\ell$ is short compared to the size of the droplet. 
Local equilibrium implies that  the volume fractions at the interface in the molecule-rich (I) and poor (II) phase  take the equilibrium values $\phi^\text{I}$ and $\phi^\text{II}$, respectively.  
In the limit of a sharp interface, the gradient of the volume fraction is aligned with the normal to the interface, and  we can write $\cos(\theta) = (\vect{n} \cdot  \nabla \phi) \, \ell/(\phi^\text{I}-\phi^\text{II})$, with the interface width $\ell$. The decay of $\phi$ around the interface is characterized by the length scale $w = \sqrt{\kappa \nu /((\chi-2)k_bT)}$~\cite{Weber_IOP_2019} and we approximate the interface width as $\ell = 3 w$, such that it is consistent with the law of Young-Dupré.
Using Eq.~\eqref{eq:bc_EQa}, we get a relationship between the contact angle $\theta$ and the coupling free energy $J$:
\begin{equation}
    \cos(\theta) = - \frac{\ell}{\kappa (\phi^\text{I}-\phi^\text{II})}
    \frac{\partial J}{\partial \phi}\, .
\label{eq:young_dupre}
\end{equation}
Accordingly, $J=0$ corresponds to a contact angle of $\pi/2$, in accordance with the law of Young-Dupré, which gives $\cos(\theta)=0$ if the surface tensions between the substrate and the molecule-poor or rich phase are identical~\cite{young1805,dupre1869, degennes1985, NJPpaper2021}. 
Thus, 
the bulk droplet makes a contact angle $\pi/2$  at the non-interacting and non-binding surface $\partial V$ (Eq.~\eqref{eq:bc_EQb}). 
Analogously, inside the surface, the boundary condition~\eqref{eq:bc_EQc} implies that surface droplets have a zero contact angle at the surface boundary $\partial S$.  

If the exchange chemical potentials between bulk and surface are not balanced, and/or one of the boundary conditions above is not fulfilled (Eqs.~\eqref{eq:equilibrium}), there will be diffusive fluxes in bulk and surface, $\vect{j}$ and $\vect{j}_s$, and a non-zero  binding flux $r$.
In this case, the system is out of  equilibrium. 
If the system is not maintained away from equilibrium~\cite{Weber_IOP_2019}, it relaxes toward thermodynamic equilibrium. 
During this relaxation, entropy is produced  until  thermodynamic equilibrium is established. In the following section, we will consider the production of entropy to derive the relationships between  the generalized fluxes and their conjugate thermodynamic forces using Onsager linear response. 

\subsection{Irreversible thermodynamics}\label{sec:IrrevThermod1}

In an isothermal system, the rate of change of the system entropy $S$ is proportional to the negative change in the total Helmholtz free energy, $T\dot S = -\dot F$~\cite{Frank_JProst2008}, where the dot indicates a total time derivative. This change in total free energy can be expressed in terms of the free energy densities by using Eq.~\eqref{eq:free_energy_F}:
\begin{align}\label{eq:total_entropy_production4}
	T \dot S &= 
	- \int_V d V \,  \partial_t \Big[ f(\phi_{}) + \frac{1}{2}\kappa \left(\nabla \phi\right)^2 \Big]\\	\nonumber
	&\quad
	- \int_S dS \, \partial_t \Big[  f_s(\phi_s) + \frac{1}{2}\kappa_s\left( \nabla_{\parallel} \phi_s\right)^2 	 +     
	 J(\phi_s, {\phi|_{z=0}})  + \frac{1}{2}\kappa_{0}(\nabla_{\parallel} \phi{|_{z=0}})^2 \Big]	\\	\nonumber
	&\quad - \int_{\partial V} dS \, \vect{n} \cdot  \vect{j}_{\text{r}1} - \int_{\partial S } dl  \, 
	\, \vect{t} \cdot  \vect{j}_{\text{r}2} \, ,
	\end{align}
where the non-dissipative free energy fluxes through the boundaries of bulk and surface are $\vect{j}_\text{r1}=\vect{j}_{}\mu_{} {/\nu}$ and $\vect{j}_\text{r2}= \vect{j}_s  \mu_s {/\nu_s}$,   respectively.

{The entropy production rate can be rewritten using the conservation laws and the boundary conditions (Eqs.~\eqref{eq:cons_laws_full_sodel}) together with the chemical potentials defined in Eqs.~\eqref{eq:chem_pot_def}:}
\begin{align}\label{eq:total_entropy_production5}
\nonumber
	T \dot S &= 
	- \int_{V } d V \, \,  \left( \nabla\mu_{}   \cdot \vect{j}_{}{/\nu} \right)
 	\\ 
 \nonumber
&\quad
- \int_S dS \,    \Big[  \nabla_{\parallel}  \mu_s  \cdot \vect{j}_s/\nu_s 
	+ (\mu_s -\mu_{}|_{z=0}) \,   {r}/\nu_s    \Big ] 
 - \int_{S} dS \, \bigg[ \mu_0 \, {\partial_t \phi|_{z=0}} /\nu_s 
 \bigg] 
 	\\ 
&\quad
 - \int_{{\partial V}} dS \, \bigg[\left(  \vect{n} \cdot \kappa_{} \left( \nabla \phi\right) \right)
{{\partial_t{\phi}}|_{\partial V}}
 \bigg] 
-  \int_{\partial S } dl  \, 
 \left(  {\vect t} \cdot \kappa_s  \nabla_{\parallel} \phi_s   \right)   \partial_t \phi_s 
 {-  \int_{\partial S } dl  \, 
 \left(  {\vect t} \cdot \kappa_0  \nabla_{\parallel} \phi|_{z=0}   \right)   \partial_t \phi|_{z=0}  } 
\, .
\end{align}

{Using irreversible thermodynamics~\cite{de2013non, livi2017nonequilibrium}, we identify the following} pairs of conjugate thermodynamic fluxes and forces:
\begin{subequations}
\begin{align}
\label{eq:j_conj}
\vect{j}_{} & \longleftrightarrow  - \nabla\mu_{}\, , &  \vect{x} & \in V \, , 
\\ 
\label{eq:js_conj}
\vect{j}_s & \longleftrightarrow  - \nabla_\parallel  \mu_s\, , &  \vect{x} & \in S\, ,
\\
\label{eq:Onsager_binding_flux}
r & \longleftrightarrow  -  \left(\mu_s -\mu|_{z=0} \right) \, ,  &  \vect{x} & \in S \, ,
\\ 
{\partial_t \phi_{}|_{z=0}} & \longleftrightarrow - \mu_0 
 \,,  &  \vect{x} & \in S \,  \, , 
\\ \label{eq:13e}
{\partial_t \phi|_{\partial V}} & \longleftrightarrow -  \left(   \vect{n} \cdot \kappa_{}  \nabla \phi_{} \right) \,,  &  \vect{x} & \in  \, \partial V  \, , 
\\ \label{eq:13f}
\partial_t \phi_s & \longleftrightarrow -  \left(    \vect{t} \cdot \kappa_s  \nabla_\parallel \phi_s \right) \,,  &  \vect{x} & \in \partial S  \,.
\end{align}
\end{subequations}
{Here, the quantities in the left column (e.g., $\vect{j}\text{}$, $\vect{j}_\text{s}$, $r$, $\partial_t \phi_{}|_{z=0}$) represent thermodynamic fluxes, whereas those in the right column (e.g., $-\nabla \mu_{}$, $-\nabla_{\parallel} {\mu}_s$, $-(\mu_s-\mu|_{z=0})$, $-\mu_0$) denote the corresponding thermodynamic forces, as introduced in Sect.~\ref{sec:Thermod1}. }
{
The fluxes in the bulk and surface, $\vect{j}$ and $\vect{j_s}$, are driven by the respective chemical potential gradients in bulk and surface, $\nabla \mu$ and $\nabla \mu_s$ (Eqs.~\eqref{eq:j_conj} and \eqref{eq:js_conj}). The binding rate $r$ results from the chemical potential difference between the surface and bulk chemical potential at to the surface, $\left(\mu_s -\mu|_{z=0} \right)$ (Eq.~\eqref{eq:Onsager_binding_flux}).
Similarly, changes in the bulk volume at the surface, $\partial_t \phi|_{z=0}$, arise due to a mismatch of the chemical potential contributions related to the coupling free energy, ${\partial J}/{ \partial \phi|_{z=0}}$, and the respective gradient free energy contributions characterized by the parameters $\kappa_0$ and $\kappa$.    
At the non-interacting and non-binding surfaces $\partial V$, the dynamics relaxes toward a neutral boundary with  $\vect{n} \cdot \kappa_{}  \nabla \phi_{}=0$. 
Similarly, at the line boundary of the binding surface, $\partial S$, the surface-bound fraction relaxes to 
$\vect{t} \cdot \kappa_s  \nabla_\parallel \phi_s=0$.}

To linear order, we obtain the following relationships between thermodynamic fluxes and forces:
\begin{subequations}\label{eq:Onsager_fluxesd}
\begin{align}
\vect{j}_\text{} &=  - \Lambda \, \nabla \mu_{}\, , &  \vect{x} & \in V  \,, \\
\vect{j}_s &=  - \Lambda_s \, \nabla_{\parallel}  {\mu}_s \, , &  \vect{x} & \in S \, ,\\
r & = - \Lambda_{r} \,  \left(\mu_s -\mu|_{z=0} \right)\,  - {\Lambda}_{{r}\kappa} \, \mu_0 
\, ,   &  \vect{x} & \in S  \, ,\\
\partial_t \phi_{}|_{z=0}&= - {\Lambda}_{{r}\kappa} \, \left(\mu_s -\mu|_{z=0} \right)  - \Lambda_{\kappa} \,  \mu_0 
\, ,  & \vect{x}  &\in S   \, , \\
\partial_t \phi|_{\partial V}&=   - \Lambda_{\kappa} \,  \left(   \vect{n} \cdot \kappa_{}  \nabla \phi_{} \right)\, , \qquad  \qquad &  \vect{x} & \in  \, \partial V  \, , \\
\partial_t \phi_s&=   - \Lambda_{\kappa_s} \,  \left(   \vect{t} \cdot \kappa_s  \nabla_\parallel \phi_s \right)\,,  \qquad \qquad &  \vect{x} & \in \partial S \, .
\end{align}
\end{subequations}
All the fluxes above 
ensure that the  entropy of the system increases when the system approaches thermodynamic equilibrium. 
In other words, linear relationship above are consistent with the second law of thermodynamics.
Here, ${\Lambda}>0$ and ${\Lambda}_{\alpha}>0$ ($\alpha=r,r{\kappa}, \kappa, \kappa_s$) denote positive kinetic coefficients, i.e., mobilities or rate coefficients. 
Specifically, $\Lambda$ and 
$\Lambda_s$ are the diffusive Onsager mobilities for bulk and surface,  and
${\Lambda}_{\rm r}$ is the  Onsager coefficient for surface binding.  Moreover, $\Lambda_{\kappa}$ and $\Lambda_{\kappa_s}$ are Onsager coefficients that govern the relaxation time toward the equilibrium boundary conditions~\eqref{eq:bc_EQa}-\eqref{eq:bc_EQc}.
Due to Onsager's reciprocal relationship and when considering linear irreversible thermodynamics, there is only one Onsager cross-coupling denoted as ${\Lambda_{r \kappa}}$.

\subsection{Dynamic equations}\label{sec:dyn_eqs}

In summary, full dynamic equations for the bulk $V$ and the binding surfaces $S$ are:
\begin{subequations}\label{eq:kinetic_eqns}
 \begin{align}
\label{eq:kinetic_eqnsb}
 \partial_t \phi_{} &= \nabla \cdot (\Lambda \nabla\mu_{})\, , & 
\vect{x}  \in V \, ,
\\
\label{eq:kinetic_eqnsa}
  \partial_t \phi_s &= \nabla_{\parallel} \cdot (\Lambda_s \nabla_{\parallel} \mu_s) + r\, ,  & \vect{x}  \in S \, ,
 \end{align}
with binding flux  given as 
\begin{align}
r & = - \Lambda_{r} \,  \left(\mu_s -\mu|_{z=0} \right)\,  - {\Lambda}_{{r\kappa}} \,  {\mu_0 } \,  .
\end{align}
{The chemical potentials depend on the two fields $\phi(\vect x, t)$ and $\phi_s(\vect{x}_\parallel, t)$ (and gradients thereof) and are given in Eqs.~\eqref{eq:chem_pot_def}. Eq.~\eqref{eq:kinetic_eqnsb} and \eqref{eq:kinetic_eqnsa} are partial differential equations of 4th order requiring two conditions at each boundary domain, i.e., at the binding surface $S$ and the non-binding and non-interacting $\partial V$, respectively, and at $\partial S$:}
\begin{align} 
\label{eq:kinetic_eqnsc}
\partial_t \phi|_{z=0}&=  - {\Lambda}_{{r}\kappa} \, \left(\mu_s -\mu|_{z=0} \right)  - \Lambda_{\kappa} \,   {\mu_0 } \, , &  \vect{x} & \in S   \, , \\
\label{eq:kinetic_eqnsf}
 - \vect{n} \cdot (\Lambda \nabla\mu_{}) &= r \frac{\nu}{\nu_s}\, ,   & \vect{x}  &\in S \, ,
 \\
 \label{eq:15f}
\partial_t \phi_{}|_{\partial V}&=   - \Lambda_{\kappa} \,  \left(   \vect{n} \cdot \kappa_{}  \nabla \phi_{} \right)\, , \qquad  \qquad & \vect{x}  &\in  \, \partial V  \, , 
\\
\label{eq:kinetic_eqnsg}
 - \vect{n} \cdot (\Lambda \nabla \mu_{}) &= 0\, ,  & \vect{x}  & \in \partial V \, ,
 \\
 \label{eq:15h}
\partial_t \phi_s&=   - \Lambda_{\kappa_s} \,  \left(   \vect{t} \cdot \kappa_s  \nabla_\parallel \phi_s \right)\,,  \qquad \qquad &  \vect{x} & \in \partial S \, ,
\end{align}
\end{subequations}
{with the second boundary condition at $\partial S$ either being periodic or $\vect{t}\cdot \nabla \mu_s=0$.
The total number of molecules $N$  is conserved during the binding dynamics between the bulk domain $V$ and $S$ which is ensured by the boundary conditions~\eqref{eq:kinetic_eqnsf}.
Please note that the right hand sides of 
Eqs.~\eqref{eq:kinetic_eqnsc},
\eqref{eq:15f} and
\eqref{eq:15h} are not sink or source terms; they describe the relaxation
toward equilibrium and cause the accumulation of molecules at the respective domain boundary.
}

At binding equilibrium $\mu_{s} = \mu|_{z=0}$ and when decoupling the surface and bulk components (i.e., $\Lambda_{r\kappa}=0$ and $\chi_{0s} = 0$), the equations above simplify to the classical Cahn-Hilliard equation (Eq.~\eqref{eq:kinetic_eqnsb}) with dynamic boundary conditions (Eq.~\eqref{eq:kinetic_eqnsc} and Eq.~\eqref{eq:kinetic_eqnsg}), as described in Refs.~\cite{Fischer_PRL1997,Fischer_1998,KENZLER2001139}.

The bulk chemical potential in bulk $\mu$ and the surface chemical potential $\mu_s$ contain the derivatives of the bulk free energy densities $f^\prime(\phi)$ and surface free energy densities $f_s^\prime(\phi_s)$ (Eqs.~\eqref{eq:chem_pot_def}), respectively, which correspond (except the multiplication with the molecular volume or molecular area) to the chemical potentials in spatially homogeneous systems. 
Such homogeneous chemical potentials can in general be expressed as follows:
\begin{subequations}
\label{eq:fprimes}
\begin{align}
    f^\prime(\phi) \, \nu & = \mu_0 + k_\text{B}T \ln \left(\gamma(\phi) \frac{\phi^{1/n}}{1-\phi}\right) \, , 
    \\
    f_s^\prime(\phi_s) \,  \nu_s & = \mu_{s,0} + k_\text{B}T \ln \left(\gamma_{s}(\phi_s) \frac{\phi_s^{1/n_s}}{1-\phi_s}\right) \, ,
\end{align}
\end{subequations}
where  $\mu_0$ and $\mu_{s,0}$ are reference chemical potentials, and $k_\text{B}$ denotes the Boltzmann constant.
Moreover, $\gamma(\phi)$ and $\gamma_{s}(\phi_s)$ are the volume and area fraction-dependent activity coefficients containing the components' interactions.

To highlight the role of such activity coefficients, 
we split up the free energies for bulk and surface, $f(\phi)=e(\phi)-s_\text{mix}(\phi)T$ and $f_s(\phi_s)=e(\phi_s)-s_{\text{mix},s}(\phi_s)T$, into the interaction free energy densities $e(\phi)$ and $e_s(\phi_s)$, and the mixing entropy densities, 
$s_\text{mix}=-(k_\text{B}/\nu) [(\phi/n) \ln \phi+ (1-\phi) \ln (1-\phi)]$ and 
$s_{\text{mix},s}=-(k_\text{B}/\nu_s) [(\phi_s/n_s) \ln \phi_s + (1-\phi_s) \ln (1-\phi_s)]$~\cite{flory1942,huggins1942,Weber_IOP_2019}. 
Here, the ratios of molecular volumes and areas between the molecule and the solvent are abbreviated by $n$ for the bulk and $n_s$ for the surface.
Performing a viral expansion of the interaction free energy densities, $e=(k_\text{B}T/\nu)[\omega \phi +\sum_{k=2} \chi(k) \phi^k]$ and $e_s=(k_\text{B}T/\nu_s) [\omega_s \phi_s +\sum_{k=2} \chi_s(k) \phi_s^k]$, 
the reference chemical potentials are $\mu_0= k_\text{B}T (\omega +n^{-1}-1)$ and $\mu_{s,0}= k_\text{B}T (\omega_s +n_s^{-1}-1)$.
Here, $\omega$ and $\omega_s$ are the bulk and surface internal free energy, and 
$\chi(k)$ and $\chi_s(k)$ are viral expansion coefficients.
Using the viral expansion,  the activity coefficients can be expressed as 
\begin{align}
\gamma(\phi) &= \exp\left(\sum_{k=2} k\, \chi(k) \phi^{k-1}\right) \, ,  \\
\gamma_s(\phi_s) &= \exp\left(\sum_{k=2} k \, \chi_s(k) \phi_s^{k-1}\right) \, , 
\end{align}
confirming that the activity coefficients depend exclusively on the components' interactions when introduced via Eqs.~\eqref{eq:fprimes}.

We have seen that the contributions $\phi^{1/n}/(1-\phi)$ and $\phi_s^{1/n_s}/(1-\phi_s)$ in Eqs.~\eqref{eq:fprimes} stem from the respective mixing entropy.
These contributions imply a  scaling of the diffusive Onsager mobilities $\Lambda$ and $\Lambda_{\rm s}$ with volume and area fractions, respectively.
This scaling can be understood when considering the dilute limits in bulk and surface ($\phi \to 0$, $\phi_s \to 0$, and $\phi \to 1$, $\phi_s \to 1$). 
In {these limits}, the activity coefficients (defined via Eqs.~\eqref{eq:fprimes}) are $\gamma=1$ and $\gamma_s=1$, and the diffusion coefficients in bulk and surface, 
$D=k_\text{B}T  \Lambda f^{\prime\prime}$ and  
$D_s=k_\text{B}T  \Lambda_s f_s^{\prime\prime}$, have to be constants, {i.e., independent of volume and area fractions.} Thus, the mixing entropy implies the following scaling for the diffusive mobilities:
\begin{subequations}
\begin{align}\label{eq:mobilities}
    \Lambda &= \Lambda_0 \, \phi(1-\phi) \, , 
    \\
       \Lambda_s &= \Lambda_{s,0} \, \phi_s(1-\phi_s) \, , 
\end{align}
\end{subequations}
where $\Lambda_{0}$
and $\Lambda_{s,0}$ are mobility coefficients that can depend on volume and area fraction.

To tailor our model for  phase separation in bulk and surface coupled via binding to a specific system,
the activity coefficients for bulk $\gamma(\phi)$
and surface $\gamma_s(\phi_s)$, have to be chosen together with the coupling free energy density $J(\phi|_{z=0}, \phi_s)$.  
{In Sect.~\ref{sect:Interaction_free_energies}, we discuss a choice of such free energy densities to study the effects of surface binding on droplet spreading.}

\section{Wetting dynamics with surface binding}\label{sect:wettingwithsurfacebinding}

In this section, we investigate how the dynamics of the spreading of a droplet in the bulk is affected by surface binding and the possibility of 
phase separation in the surface. In this section, we consider a two-dimensional system with a one-dimensional boundary; see Appendix~\ref{APP:3D_cylinder} and Fig.~\ref{fig:figure3b_comparison} for the results considering a three-dimensional system with axial symmetry. using 2D numerical simulations. 
For such studies, we set for simplicity ${\Lambda_{r \kappa} = 0}$ and $\Lambda_{\kappa} = \Lambda_{\kappa_s} = \infty$, implying that the equilibrium boundary conditions Eqs.~\eqref{eq:bc_EQa}-\eqref{eq:bc_EQc} hold during the spreading dynamics.
Moreover, we consider the mobility coefficients, $\Lambda_0$ and $\Lambda_{s,0}$, {in Eq.~\eqref{eq:mobilities}} to be constants, {and the free energy cost $\kappa_0 = 0$, for simplicity.}

\subsection{Interaction free energies}\label{sect:Interaction_free_energies}

To study spreading, we consider Flory-Huggins free energy densities for bulk and surface,
\begin{subequations}
\begin{align}
\label{eq:fbulk}
  f(\phi) &= \frac{k_\text{B} T}{\nu} \Big[ \frac{1}{n} \phi\, \ln \phi  +  (1-\phi)\, \ln \left(1-\phi\right) - \chi \phi^2 + \omega \phi
    \Big]\, ,\\ 
\label{eq:fsurface}
       f_s(\phi_s) &= \frac{k_\text{B} T}{ \nu_s} \Big[ \frac{1}{n_s} \phi_s\, \ln \phi_s + (1-\phi_s)\, \ln(1-\phi_s) - \chi_s \phi_s^2
       + \omega_s \phi_s \Big]\, .
\end{align}
Using Eq.~\eqref{eq:fprimes}, the two free energies correspond to the  activity coefficients, respectively: 
\begin{align}
    \gamma(\phi)&= \exp\left(-2\chi \phi\right) \, , \\
    \gamma_s(\phi_s)&= \exp\left(-2\chi_s \phi_s\right) 
    \, .
\end{align}
In other words, we consider a mean-field free energy up to the second order, with the  coefficients in the viral expansion $\chi(2)=-\chi$ and $\chi_s(2)=-\chi_s$.

Interactions between the surface and the bulk are captured by the coupling free energy density~\cite{NJPpaper2021}:
\begin{align}\label{eq:coupling_free_energy}
J(\phi|_{z=0}, \phi_s) & = \frac{k_\text{B} T}{\nu_s} \Big[
\omega_0 \, \phi|_{z=0} + {\chi}_{00} \, \phi|_{z=0}^2 + \chi_{0s} \, \phi|_{z=0} \, \phi_s \Big] \, ,
\end{align}
\end{subequations}
which encompasses all relevant terms up to the second order. Note that a term proportional to $\phi_s^2$ already exists in surface free energy density $f_s$ (Eq.~\eqref{eq:fsurface}). 
In Eq.~\eqref{eq:coupling_free_energy}, the parameter $\omega_0$ represents the internal free energy of a bulk molecule at the surface. 
When the surface is attractive for bulk molecules, $\omega_0<0$. The coefficient $\chi_{00}$ quantifies the interactions among bulk molecules at the surface
leading to enrichment ($\chi_{00}<0$) or depletion ($\chi_{00}>0$)
at the surface. For simplicity, we have set $\chi_{00}$ to zero for the studies shown in this section. Furthermore, $\chi_{0s}$  describes the interactions between bound and unbound molecules at the surface. 
For all our studies, we have assigned a negative value to $\chi_{0s}$. 
This choice corresponds to the case that molecules, bound or unbound, attract each other. 
For a comprehensive thermodynamic study of all three parameters, we refer the reader to Ref.~\cite{NJPpaper2021}.

\subsection{Non-dimensionallization of dynamic equations and dimensionless parameters}

To solve the dynamics equations, we write them in a dimension-less form. 
We obtain non-dimensional dynamic equations by rescaling length and  time scales  as follows:
\begin{subequations}
\begin{align}
   &\vect{x} \to \vect{x}  \, \nu^{1/3} \, , 
   \\
   &t \to t \, \nu^{2/3}/(\Lambda_0 k_\text{B}T )  \, ,\\
&\tilde{V} = \{ \vect{x}/\nu^{1/3}   \, \, \,  | \, \vect{x} \in V\} \, , \\
&\tilde{S} = \{  \vect{x}_\parallel/\nu^{1/3}   \, | \,\vect{x}_\parallel \in S\} \, , 
\end{align}
\end{subequations}
where $\tilde{V}$ and $\tilde{S}$ are the rescaled bulk and surface. 
This choice leads to the  following non-dimensional parameters:
\begin{subequations}
\label{eq:nondim_parameters}
\begin{align}
\mathcal{D}_{\rm s} &=\frac{\Lambda_{s}}{\Lambda} \, , \\
k_{ r} &= {\Lambda}_{r} \frac{\nu^{2/3}}{\Lambda_0}\, .
\end{align}
\end{subequations}
Moreover, we introduce the following rescaled quantities: 
the rescaled bulk and surface free energy densities, $\tilde{f}(\phi)=({\nu}/{k_{\rm B}T})f(\phi)$
and  $\tilde{f}_{s}(\phi_{ s})=(\nu/k_{\rm B}T)f_{s}(\phi_{s})$,
the rescaled coupling free energy density
$\tilde{J}(\phi|_{z=0}, \phi_s) = (\nu_s/k_{\text{B}}T)J(\phi_s, \phi|_{z=0}) $,
and the rescaled coefficients characterizing the free energy costs for gradients, $\tilde{\kappa} =\kappa  \nu^{1/3}/({k_{\text{B}}T})$, and
$\tilde{\kappa}_{\rm s} =  (\nu_{s}/(k_{\text{B}}T\nu^{2/3}) \kappa_{s} $.
The dimensionless  equations governing the kinetics of the system are thus given as:
\begin{subequations}
\label{eq:kinetic_eqns_non_dim}
\begin{align} 
\partial_t \phi &= \nabla \cdot \left[  \phi_{} (1 - \phi_{})  \left( 
\tilde{f}^{\prime\prime} \, 
\nabla \phi_{}   -  \tilde{\kappa}  \nabla^3 \phi   \right)   \right] \, ,  \quad & \vect{x}  &\in \tilde{V} \, , \\
\partial_t \phi_{s} &= \nabla_\parallel \cdot \left[ \mathcal{D}_{s} \phi_s \left(1 - \phi_s \right) \left( 
\tilde{f}_{s}^{\prime\prime} \, 
\nabla_\parallel \phi_{s}
+ \nabla_\parallel \frac{\partial \tilde{J}}{\partial \phi_{s}}  -  \tilde{\kappa}_{s}  \nabla_\parallel^3 \phi_{s}   \right )     \right ]  + \tilde{r}(\phi_{s}, \phi|_{z=0})  \, ,  \quad & \vect{x}  &\in 
\tilde{S} \, .
\end{align}
The dimensionless binding flux reads:
\begin{equation}
    \tilde{r}  = - k_{r} \left(\frac{\mu_{s}}{k_{\rm B}T} - \frac{\mu|_{z=0}}{k_{\rm B}T}\right)\, ,
\end{equation}
with the chemical potential in bulk and surface,
\begin{equation}
    \frac{\mu_{s}}{k_{\rm B}T} = \frac{\partial \tilde{f}_{s}}{\partial \phi_{s}}
+ \frac{\partial \tilde{J}}{\partial \phi_{s}}  -  \tilde{\kappa}_{s} \nabla_\parallel^2 \phi_{s}\,, \qquad
    \frac{\mu}{k_{\rm B}T}  = \frac{\partial \tilde{f}}{\partial \phi} -  \tilde{\kappa} \nabla^2 \phi    \, .
\end{equation}
The boundary conditions in a dimensionless form are: 
\begin{align} 
0&= \frac{\partial \tilde{J}}{ \partial \phi|_{0}} +  \frac{\nu_{s}}{\nu^{2/3}}\tilde{\kappa} \, \vect{n} \cdot  \left( \nabla \phi_{}\right)|_0 \, , & \vect{x}  &\in \tilde{S}   \, , \\
0 &=    \tilde{\kappa} \, \vect{n} \cdot   \nabla \phi \, , \qquad  \qquad  & \vect{x}  &\in  \, \partial \tilde{V}  \, , 
\\
0 &= \tilde{\kappa}_{s} \, \vect{t} \cdot   \nabla_\parallel \phi_{s} \,,  \qquad \qquad  & \vect{x}  &\in \partial \tilde{S},
\\
 -    \frac{\nu^{2/3}}{\nu_s } \, \tilde{r} & =  \phi (1 - \phi) \,  \vect{n} \cdot \nabla \frac{\mu_{{}}}{k_{\text{B}}T} \, ,  \quad & \vect{x}  &\in \tilde{S}  \, , 
 \\
 0 & = \phi (1 - \phi) \,  \vect{n} \cdot \nabla \frac{\mu_{{}}}{k_{\text{B}}T}  \, ,  \quad & \vect{x}  &\in \partial \tilde{V} \, .
\end{align}
\end{subequations}

To numerically solve the dynamic equations~\eqref{eq:kinetic_eqns_non_dim}, we develop a numerical solver for the corresponding systems of partial differential equations. Specifically, we employ the finite difference method for spatial discretization and the Crank-Nicolson scheme combined with the energy quadratization method~\cite{Jia_XF_YZ_XP_XG_Jun_Q2018,ZYLW_SISC_2016,Zhao_W2018_2} to discretize time. 

\subsection{Spreading  kinetics towards complete wetting}
\label{sect:spreading_toward_complete}

We use our dynamic theory of  wetting to investigate the influence of surface binding on the kinetics toward a completely wetted state. 
To this end, the system is initialized with a droplet  near the surface where no molecules are initially bound (Fig.~\ref{fig:sketch}b,c).
This initial droplet has local volume fractions corresponding to the equilibrium phase diagram. Such phase diagrams are, in general, determined by the interaction parameters $\chi$, $\chi_s$, and $\chi_{0s}$. For simplicity, we choose equal molecular volumes and areas between molecule and solvent in bulk and surface, i.e., $n=1$ and $n_s=1$.  
We also fix $\chi_s = \chi = 2.5$ and $\chi_{0s} = -0.5$, and $N\nu/|V|={0.17}$ for the forthcoming studies ensuring that droplets are thermodynamically stable in bulk and surface; see Appendix~\ref{APP:phase_diag} for details.
Moreover, refer to Table~\ref{tab:model_value} for a summary of the chosen parameter values for our wetting studies.
For the sake of simplicity, the surface internal free energy $\omega_s$ has been assigned a value of zero. 
To describe 
repulsive interactions between the surface and bulk molecules, 
we choose $\omega_0 = 0.17$ in Eq.~\eqref{eq:coupling_free_energy}, and we choose the interaction parameter of bulk molecules at the surface as $\chi_{00} = 0$. 

To illustrate the qualitative behavior of the contact angle $\theta$,
we consider the wetting boundary condition 
Eq.~\eqref{eq:young_dupre}, which is valid when the triple line is at local equilibrium. For the parameter choices discussed above, this condition gives $\cos{(\theta)}\propto -\partial J/\partial \phi= -(k_\text{B}T/\nu_s) [\omega_0 +\chi_{0s}\phi_s]$.
We see that the contact angle varies with the area fraction of bound molecules. 
For our choices $\omega_0 >0$ (repulsive) and $\chi_{0s}<0$ (attractive),  $\cos{(\theta)}(\phi_s=0)<0$, corresponding to partial wetting with $\pi/2<\theta<\pi$ (for $\omega_0 = 0.17$), or the case of dewetting for even larger values of $\omega_0$. When more molecules are bound to the surface ($\phi_s>0$), $\cos{(\theta)}(\phi_s)$ increases linearly with a steeper slope for larger  values of the surface-bulk coupling, $|\chi_{0s}|$.  For large enough values of $|\chi_{0s}|$ and $\phi_s$, the contact angle $\theta=0$ corresponding to complete wetting. 
In summary, when more molecules are bound to the surface, the considered system tends to cross from partial wetting to complete wetting. 
As we initialize the system without molecules bound to the surface, we expect such a trend in wetting behavior upon surface binding. 

For simplicity, we consider a two-dimensional domain for the bulk with a one-dimensional surface; see green line Fig.~\ref{fig:figure2a}a. 
In addition to the boundary conditions given in Eqs.~\eqref{eq:kinetic_eqns_non_dim}, 
periodic boundary conditions are applied along the left and right boundaries at $\partial S$ and $\partial V$. An exception is the $\partial V$ boundary opposite to the binding surface $S$ where we apply no flux boundaries (non-binding), as stated in Eq.~\eqref{eq:bc_flux_nobinding}.

\subsubsection{Surface binding nucleates a surface droplet that accelerates  spreading}

\begin{figure*}[t]
\centering
{\includegraphics[height=15cm]{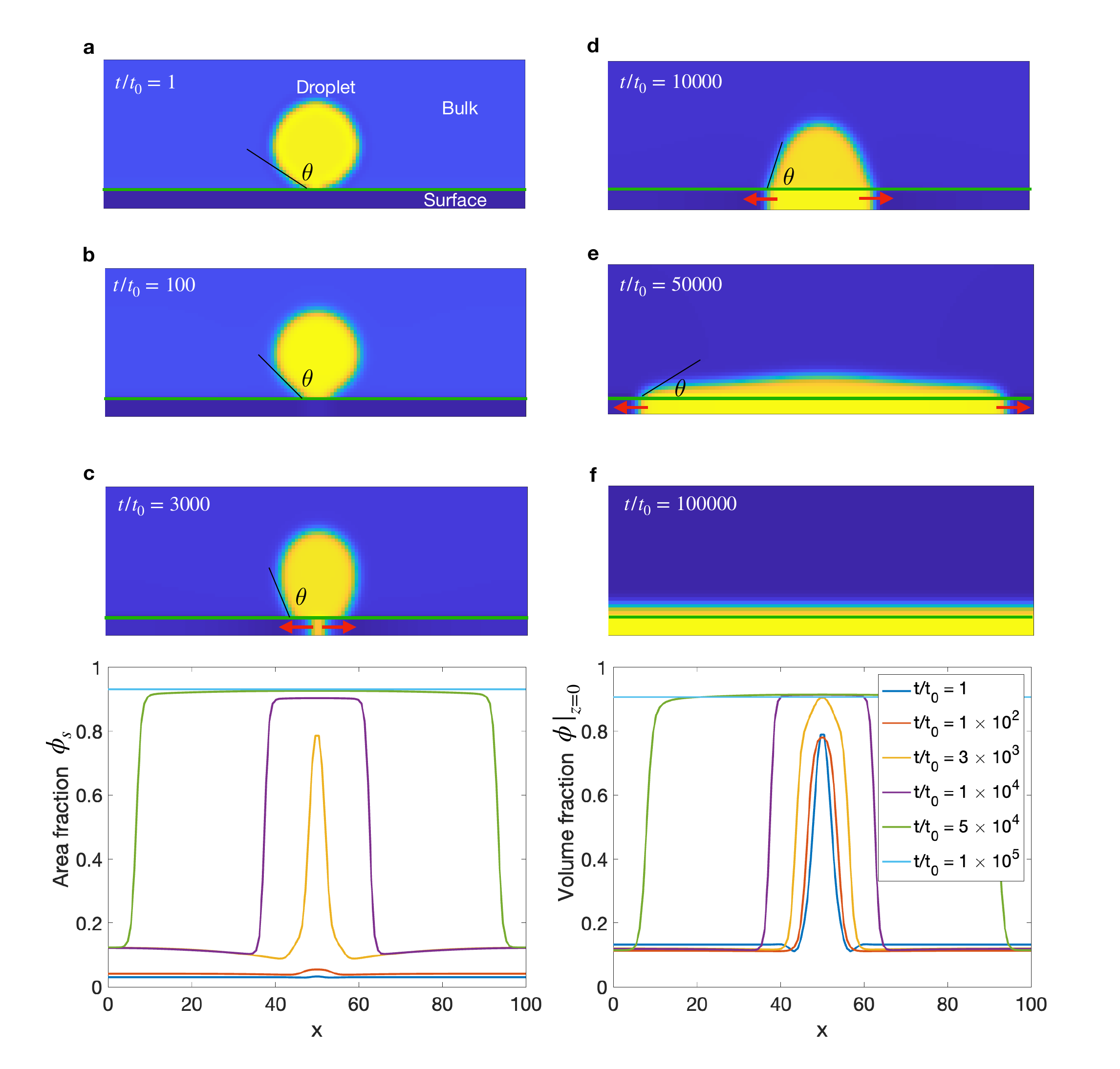}}
\caption{\textbf{Bulk droplet spreading on a binding surface toward complete wetting.} 
\textbf{a)} Initially, the bulk droplet forms a bridge, and 
\textbf{b)} slowly spreads on the surface (green line).
\textbf{c,d)} After nucleation of a surface droplet underneath the bulk droplet,
\textbf{e)} both droplet interfaces move synchronously, quickly covering the surface and 
finally leading to a completely wetted state \textbf{f)}. $\theta = 0^{\text{o}}$ at equilibrium. See Appendix \ref{APP:phase_diag} for details.
{\textbf{g,h)} show the surface-bound area fraction $\phi_s$ and boundary volume fraction of $\phi$ at $z=0$, i.e. $\phi|_{z=0}$, of the snapshots in plot \textbf{(a-f)}. }
Parameters: $k_r = 10^{-4}$, $\chi_{0s} = -0.5$, $t_0= \nu^{2/3}/(\Lambda_0 k_\text{B}T)$, more see Table~\ref{tab:model_value}. 
{For all the numerical simulations in this study, we use $\Delta t = 10^{-1}$ and $\Delta x = 1/128$ as the temporal and spatial mesh size. }
}
\label{fig:figure2a}
\end{figure*}

After initializing the droplet adjacent to the surface, it develops a small bridge of molecule-rich phase with the surface; Fig.~\ref{fig:figure2a}a.
As time progresses, the droplet slowly wets on the surface; Fig.~\ref{fig:figure2a}b. 
The dynamics {is} slow because the surface without binding is weakly repulsive ($\omega_0>0$). 
In the absence of binding, the droplet approaches a partially wetted state with a contact angle $\theta>\pi/2$. 
Note that without binding ($\phi_s=0$), $\omega_0=0$ and $\chi_{00}=0$ in Eq.~\eqref{eq:coupling_free_energy} leads to $J=0$ and a contact angle of $\theta=\pi/2$ (Eq.~\eqref{eq:young_dupre}).  
However, molecules additionally start accumulating in the surface by binding.
This accumulation is more pronounced right underneath the bulk droplet (Fig.~\ref{fig:phi_s_t}a in Appendix~\ref{APP:nucleation_surface_drop}).
Once this local volume fraction exceeds the saturation volume fraction, a droplet gets nucleated inside the surface (Fig.~\ref{fig:figure2a}c).
This surface droplet grows quickly due to the influx of droplet material from the bulk droplet right above, indicated by the red arrow in the figure. 
As a result, the interface in the surface and the bulk  coincide and start moving together (Fig.~\ref{fig:figure2a}d).
After that, the bulk droplet spreads quickly until it completely wets the surface (Fig.~\ref{fig:figure2a}e,f).
Concomitantly, the area fraction far away from the surface droplet reaches the equilibrium value from below (Fig.~\ref{fig:phi_s_t}b in Appendix~\ref{APP:nucleation_surface_drop}).

\begin{figure*}[t]
\centering
{\includegraphics[height=12cm]{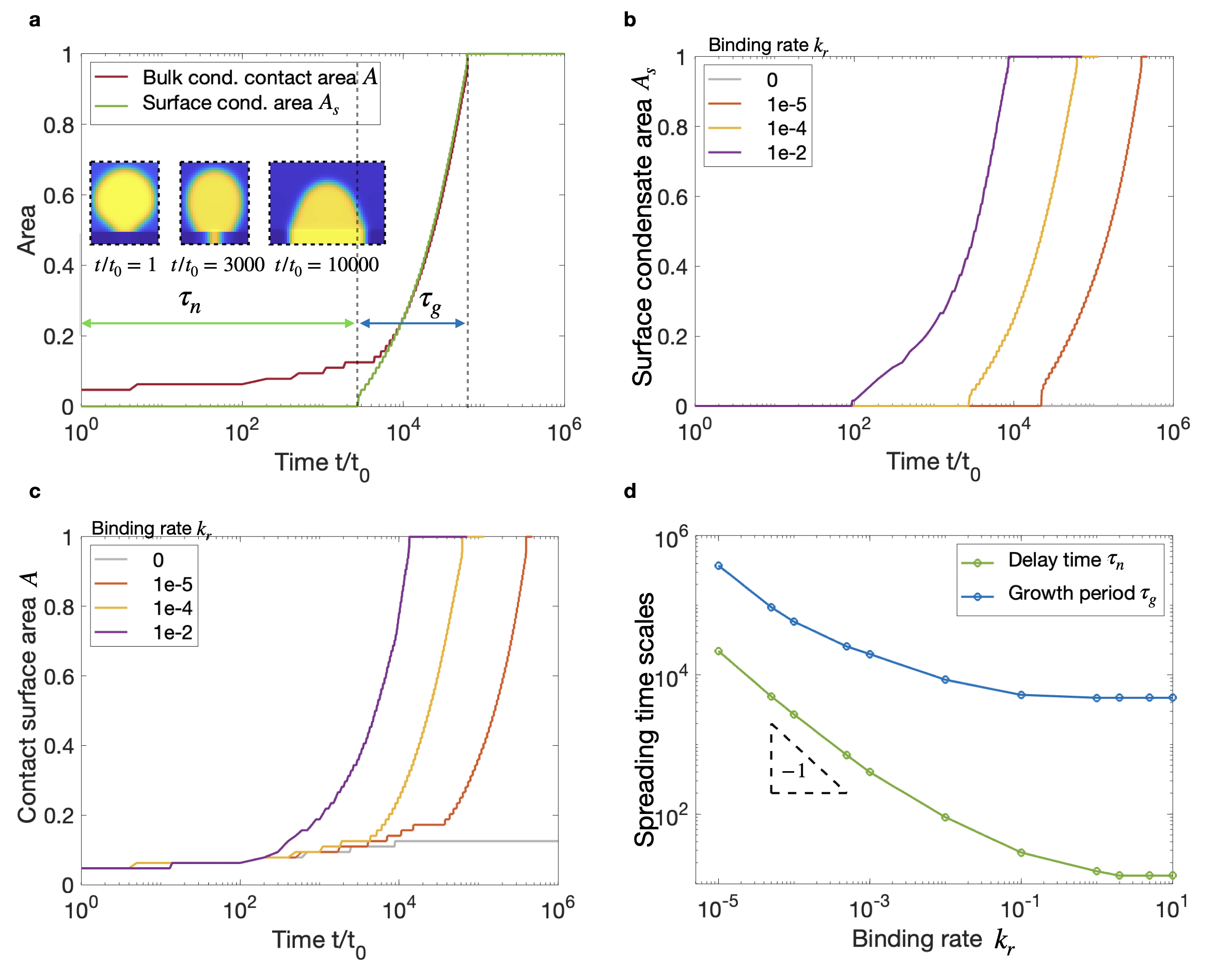}}
\caption{\textbf{Spreading dynamics toward a completely wetted state is accelerated by surface binding.} 
\textbf{a)} The contact area of the bulk droplet $A$ initially increases slowly while speeding up after the nucleation of a surface droplet around $t/t_0=3\cdot 10^3$, evidenced by the fast increase of the surface droplet area $A_s$. 
The time once $A_s$ is non-zero corresponds to the nucleation time $\tau_{\rm n}$ of a surface droplet, and $\tau_{\rm g}$ is the period for it to grow until it covers the full surface.
\textbf{b)} The area of the surface droplet $A_s$ shows that the nucleation time $\tau_{\rm n}$ decreases with increasing rescaled binding rate $k_r$. 
\textbf{c)} The acceleration in the growth of the bulk droplet area shifts toward earlier times for increasing values of $k_r$.
\textbf{d)} The growth period $\tau_{\rm g}$ and the nucleation time $\tau_{\rm n}$ both scale proportional to $k_r^{-1}$ for increasing rescaled binding rate $k_r$. For large values, diffusion in bulk and surface becomes rate-limiting, leading to a plateau.  
Parameters: $t_0= \nu^{2/3}/(\Lambda_0 k_\text{B}T)$, more see Table~\ref{tab:model_value}.
} 
\label{fig:figure2b}
\end{figure*}

To characterize how the bulk droplet affects the lower dimensional surface droplet that had been nucleated via binding, we determine the contact area of the bulk droplet $A$ on the surface and the area of the surface droplet $A_s$ with time.  
As illustrated in Fig.~\ref{fig:figure2a}a-f, 
the area of the bulk droplet $A$ first grows slowly. 
Suddenly, its growth speeds up quickly, leading to a completely wetted surface (Fig.~\ref{fig:figure2b}a). 
The time at which the growth of the bulk droplet speeds up coincides with the time $\tau_{\rm n}$
when a surface droplet is nucleated underneath the bulk droplet. 
Moreover, the time of fast spreading of the bulk droplet corresponds to the growth time $\tau_{\rm g}$ of the surface droplet underneath, i.e., the time it takes for the surface droplet area $A_s$ to grow from zero to full surface coverage.

Increasing the binding rate $k_r$ to the surface  speeds up both, nucleation and growth of surface droplets. This is evident by the shift of $A$ and $A_s$ to smaller times (Fig.~\ref{fig:figure2b}b,c). The nucleation time $\tau_{\rm n}$ and the growth time $\tau_{\rm g}$ decays algebraically for smaller binding rate $k_r$, while both saturate for very large values of $k_r$. The saturation of both processes results from diffusion in bulk and membrane becoming rate limiting.

\subsubsection{Incompatible phase equilibrium in bulk and surface}

In our studies, the surface droplet grows until the surface is homogeneously covered by bound molecules.  
The absence of phase separation in the surface results from  phase equilibria in bulk and surface being in general incompatible, except for very special parameter choices. 
Incompatible means that the condition for phase equilibrium in bulk ($\mu^\text{I}=\mu^\text{II}$) and surface ($\mu_s^\text{I}=\mu_s^\text{II}$) cannot be satisfied concomitantly. 

\begin{figure*}[t]
\centering
{\includegraphics[width=16cm]{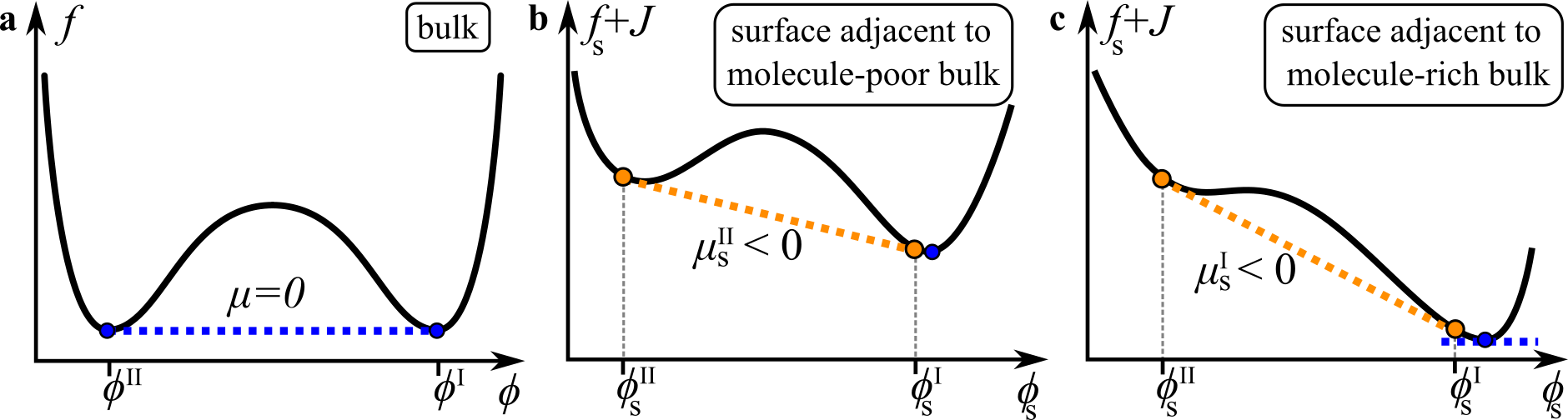}}
\caption{\textbf{Incompatible phase equilibria in bulk and surface} 
\textbf{a)} The bulk free energy density $f_s$, where the Maxwell construction of the symmetric double well potential corresponds to a bulk chemical potential $\mu=0$. 
\textbf{b)} The total free energy density of the surface adjacent to a molecule-poor bulk phase, and \textbf{c)} adjacent to a molecule-rich bulk phase. 
The Maxwell construction gives two different pairs of equilibrium area fractions, $\phi_{s}^\text{I}$ and $\phi_{s}^\text{II}$. 
However, both cases are incompatible with phase equilibrium in the bulk, \textbf{a}. In other words, thermodynamic equilibrium is a homogeneous state corresponding to the global minimum in \textbf{c}.
}
\label{fig:schematic}
\end{figure*}

To understand such incompatible equilibria for our coupled bulk-surface system, we first take a closer look at the free energy densities in bulk and surface, $f$ and $f_s$; see Fig.~\ref{fig:schematic}a for a schematic sketch of $f$. For equal molecular volumes of molecule and solvent in the bulk ($n=1$),  the free energy density in the bulk $f$ is a symmetric double-well potential (Eq.~\eqref{eq:fbulk}), where $\phi^\text{I}$ and $\phi^\text{II}$ are the equilibrium  volume fractions in the molecule-rich and poor phase.
In this case, the slope of the Maxwell construction corresponds to the bulk chemical potential $\mu=0$.
Though the surface free energy density $f_s$ is also symmetric for $n_s=1$ (Eq.~\eqref{eq:fsurface}), 
the total free energy density  of the surface, $f_s+J$, is not symmetric when the interaction parameters $\omega_0$, $\chi_{00}$ and $\chi_{0s}$ are non-zero in Eq.~\eqref{eq:coupling_free_energy}. 
Since the coupling free energy density $J(\phi|_{z=0}^\text{I/II}, \phi_s)$ depends on the bulk phases right above, 
the total free energy density of the surface $(f_s+J)$ is different below the molecule-rich or molecule-poor phase, respectively (Fig.~\ref{fig:schematic}b,c). Since the coupling $\chi_{0s}<0$ is attractive, the Maxwell construction for phase coexistence in the surface would require that the surface chemical potential adjacent to the molecule-poor bulk phase, $\mu_s^\text{II}<0$. Since $\phi|_{z=0}^\text{I}>\phi|_{z=0}^\text{II}$, the Maxwell's slope $\mu_s^\text{I}<0$ is even more negative.
As thermodynamic equilibrium also requires that binding equilibrium is satisfied (Eq.~\eqref{eq:equilibrium}), phase coexistence in bulk and surface is only possible if $\mu_s^\text{I}=\mu^\text{I}=0$ and $\mu_s^\text{II}=\mu^\text{II}=0$. These conditions cannot be satisfied in general; the only exception is when all three interaction parameters $\omega_0$, $\chi_{00}$ and $\chi_{0s}$ vanish. 
If not, a bulk droplet can only coexist with a  surface homogeneously covered by bound molecules.
The equilibrium surface area fraction corresponds to the global minimum in the total surface free energy densities $(f_s+J)$ which is the surface molecule-rich phase in Fig.~\ref{fig:schematic}c (indicated by blue line).

\subsubsection{Scaling of the nucleation time $\tau_{\rm n}$}

The droplet phase is nucleated in the surface directly underneath the center position of the bulk droplet, $\vect{x}_{\parallel}|_\text{center}$. Due to the mirror symmetry at this center position, lateral gradients have to vanish at the center in the bulk and surface ($\nabla_\parallel \phi_s =0$ and $\nabla_\parallel \phi =0$ at 
$\vect{x}_{\parallel}|_\text{center}$).
Consequently, the dynamic equation for the area fraction $\phi_{s}$ at the center  positions reads
\begin{equation}
\partial_t\phi_{s}\big|_{\rm center} = k_{r}(\mu-\mu_{\rm s})\big|_{\rm center}.
\end{equation}
Nucleation in the surface induced by the bulk droplet corresponds to the scenario where $\mu>\mu_{\rm s}$ (Fig.~\ref{fig:schematic}a-c), i.e., molecules bind to the surface and nucleate a surface droplet when the local area fraction exceeds the equilibrium concentration (Appendix~\ref{APP:phase_diag}). 
If bulk diffusion is fast compared to binding ($k_r \ll 1$ in Eq.~\eqref{eq:nondim_parameters}), the bulk volume fractions inside and outside of the droplet are close to their respective equilibrium value at all times. This corresponds to  a spatially constant $\mu$ at the moment of nucleation. This also implies that $\mu_s$ is spatially constant on the surface.  
Thus, the time to form a nucleus on the surface right beneath the center of the bulk droplet scales,  $\tau_{\rm n} \sim \left( \partial_t\phi_{s}\big|_{\rm center} \right)^{-1} \simeq k_{r}^{-1}$. This scaling agrees with the result obtained from numerically solving the dynamics equations~\eqref{eq:kinetic_eqns_non_dim}; see Fig.~\ref{fig:figure2b}d. 

For fast binding compared to bulk diffusion ($k_{r}\gg 1$), nucleation of surface droplets is limited by diffusion in surface and bulk, while the binding is at equilibrium at all times, $\mu_s \simeq \mu$. Thus, the nucleation time $\tau_{\rm n}$ becomes constant and independent of the binding rate $k_r$, which is consistent with the results in Fig.~\ref{fig:figure2b}d.

\subsubsection{Scaling of the growth time $\tau_{\rm g}$ for complete wetting}\label{sect:taugcomplete}

Now, we derive the scaling behavior of the growth time $\tau_{\rm g}$ with the rescaled binding rate $k_r$.
To this end, we first note that in the
surface region underneath the droplet, the binding flux is negligible if the surface area fraction attains a value that correspond to a local chemical potential $\mu_{s}^\text{I}=0$. 
A sizable binding flux is hence limited to the surface region adjacent to the molecule-poor bulk phase, where we approximate the binding flux as $r=k_{r}\mu_{s}^\text{II}$. 
In the following, we denote 
the position of the interface of the surface droplet by $X_0$. 
We note that in the case of complete wetting, the position of the droplet interface coincides with $X_0$ throughout the late stage of the spreading process (Fig.~\ref{fig:figure2a}d-f). 
The time evolution of the average area fraction $\bar{\phi}_{s}$  (Eq.~\eqref{eq:average_area_fraction}) in rescaled units thus reads
\begin{equation}
    \frac{d\bar{\phi}_{\rm s}}{dt} = -2k_{r}\frac{\mu_{s}^\text{II}}{k_\text{B}T}\left(1-\frac{X_0}{X_{\rm max}}\right)\, ,
\label{eq:number_of_particles_3}
\end{equation}
where $X_{\rm max}=L/2$. 
In addition, volume conservation inside the surface implies: \\ $\bar{\phi}_{s}=2\left(\phi_{\rm s}^\text{I}X_0 + \phi_{s}^\text{II}(X_{\rm max}-X_0)\right)/X_{\rm max}$. Taking the time derivative gives 
\begin{equation}
    \frac{d\bar{\phi}_{s}}{dt} = 2\frac{\phi_{s}^\text{I}-\phi_{s}^\text{II}}{X_{\rm max}}\frac{dX_0}{dt}.
    \label{eq:number_of_particles_1}
\end{equation}
The interface speed of the surface droplet, $dX_0/dt$, is 
\begin{equation}
    \frac{dX_0}{dt}\simeq -k_{\rm r} \frac{\mu_{s}^\text{II}}{k_\text{B}T} \left(X_{\rm max}-X_0\right),
\end{equation}
which implies that the growth  time  scales  as $\tau_{\rm g} \sim k_{r}^{-1}$.

\subsection{Spreading  kinetics towards partial wetting}
\label{sect:spreading_toward_partial}

Now, we use our theory of  wetting to investigate the influence of surface binding on the kinetics towards a partially wetted state. 
To this end, the system is initialized similarly as described in Sect.~\ref{sect:spreading_toward_complete}, i.e., with a droplet near the surface at which no molecules are initially bound (Fig.~\ref{fig:sketch}b,c).

\subsubsection{Surface Binding Controls the Droplet Partial Wetting Kinetics}\label{subsec:partial_wetting}

\begin{figure*}[b]
\centering
{\includegraphics[height=15cm]{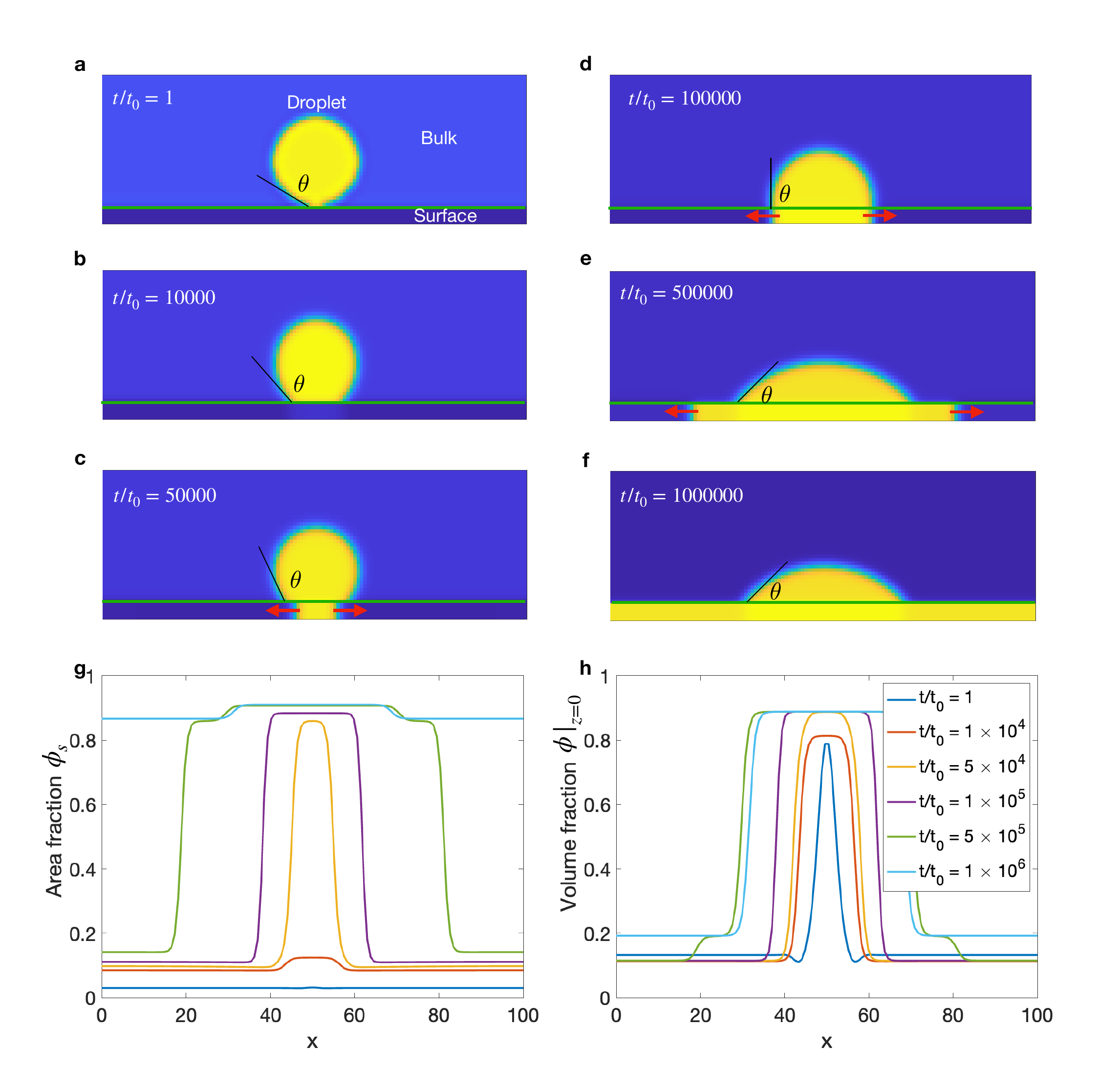}}
\caption{\textbf{Bulk droplet spreading on a binding surface towards partial wetting.} 
\textbf{a)} After bridge formation,  and 
\textbf{b)} slow spreading on the surface (green line),
\textbf{c)} a surface droplet is nucleated. 
\textbf{d,e)} The surface droplet grows and overtakes the interface of the bulk droplet, 
which arrests a constant contact angle $\theta$.
\textbf{f)} While the surface droplet grows towards a state that completely covers the surface, the bulk droplet shrinks a bit since the bulk loses molecules that bind to the surface. 
{\textbf{g,h)} show the surface-bound area fraction $\phi_s$ and boundary volume fraction of $\phi$ at $z=0$, i.e. $\phi|_{z=0}$, of the snapshots in plot \textbf{(a-f)}. }
Parameters: $k_r = 10^{-5}$, $\chi_{0s} = -0.33$, $t_0= \nu^{2/3}/(\Lambda_0 k_\text{B}T)$, more see Table~\ref{tab:model_value}.
{For all the numerical simulations in this study, we use $\Delta t = 10^{-1}$ and $\Delta x = L/128$ as the temporal and spatial mesh size. }
} 
\label{fig:figure3a}
\end{figure*}

To study the effects of binding on the spreading of a bulk droplet toward a partially wetted state, we reduced the attractive coupling between the bulk and surface, $\chi_{0s}$,   relative to the previous study towards complete wetting. Specifically, we choose $\chi_{0s} = -0.33$. The other parameters and boundary conditions remained the same as in the earlier study; see Table~\ref{tab:model_value}. 

The early dynamics is similar to spreading towards a completely wetted state (Fig.~\ref{fig:figure3a}a-d): A bridge of dense phase forms from the bulk droplet, which spreads on the surface towards contact angle $\theta<\pi/2$. 
This is followed by a nucleation of a droplet in the surface which quickly grows due to binding from the bulk droplet.
Due to the still attractive coupling interactive interactions $\chi_{00}$, the contact angle of the bulk droplet and the surface increases. 

The first difference to the case of complete wetting studied in Sect.~\ref{sect:spreading_toward_complete}
occurs when the interface of the surface droplet aligns with the interface of the bulk droplet. 
Owing to the less attractive bulk-surface interaction, $\chi_{0s}$, the bulk droplet can only partially wet the dense phase of the surface droplet.
In fact, for time scales larger than the meeting time of both interfaces, $t_o$, the contact angle $\theta$ remains stationary and finite (Fig.~\ref{fig:figure3a}d-f).
For $t>\tau_{\rm o}$, the interface of the surface drop overtakes the interface of the bulk droplet. 
There is even a negative feedback: As the surface droplet approaches full surface coverage, the bulk droplet shrink at approximately fixed contact angle.
This feedback arises because binding to the surface lowers the average volume fraction in the bulk. 
This effect vanishes when the bulk contains much more molecules than the surface, i.e., for macroscopic systems. However, such effect could be relevant for small system such as cellular compartments and small reaction containers used in system chemistry.

\begin{figure*}[h!]
\centering
{\includegraphics[height=12cm]{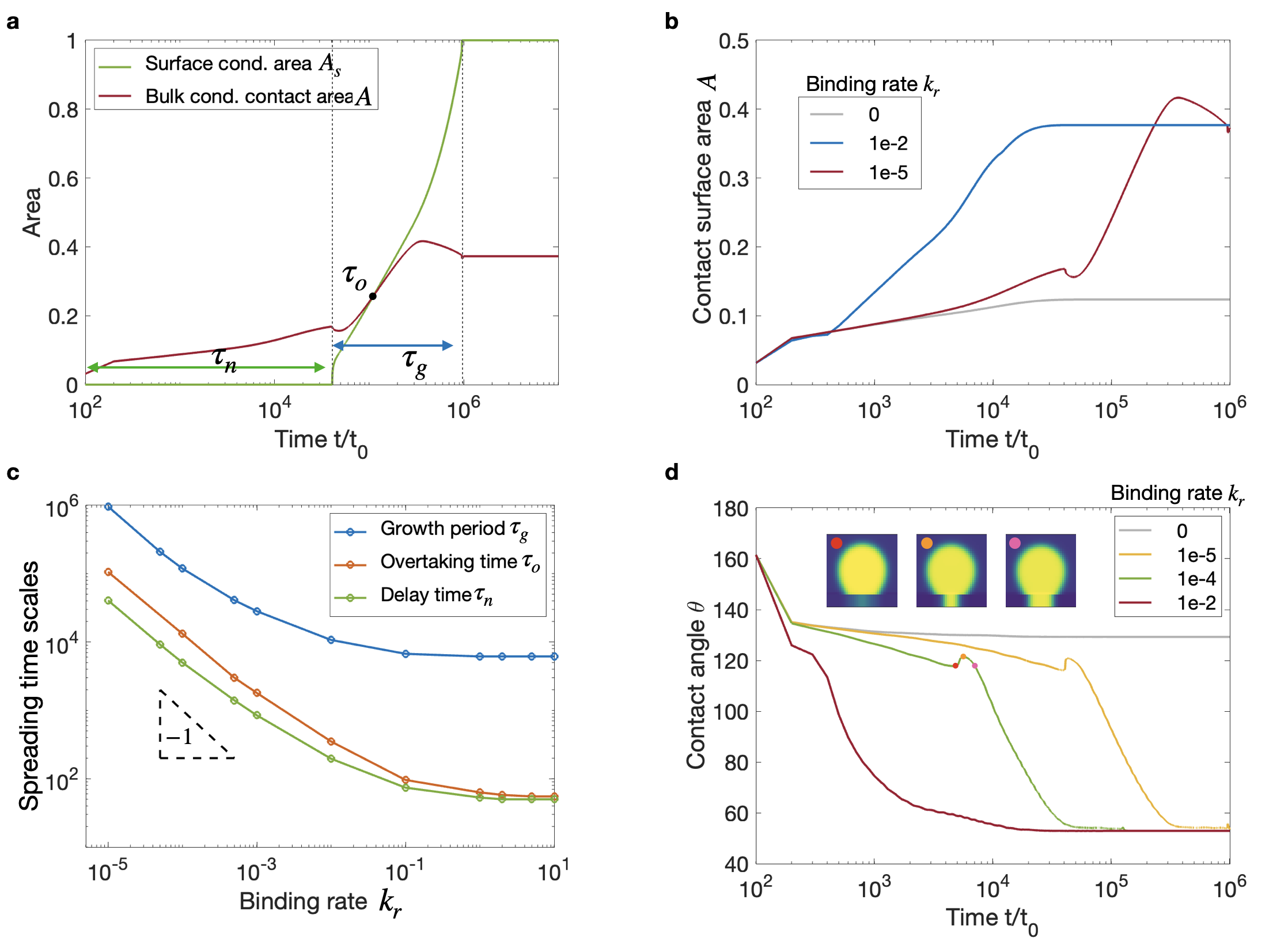}}
\caption{
\textbf{Spreading dynamics toward a partially wetted state is accelerated by surface binding.} 
\textbf{a)} The contact area of the bulk droplet A initially increases slowly until a surface
droplet is nucleated at $t/t_0 = \tau_{\rm n}$. The interface of the surface droplet takes over the interface of
the bulk droplet at time $\tau_{\rm o}$ and then quickly covers the full surface with a total growth time $\tau_g$.
In contrast, the bulk droplet stops spreading, leading to a constant contact angle $\theta$ and a slight
shrinkage in drop volume due to the loss of bulk molecules by binding.
\textbf{b)} The area of the bulk droplet $A$ shows that the spreading dynamics of the bulk droplet are more accelerated for faster rescaled binding rate $k_r$. 
\textbf{c)} The growth period $\tau_{\rm g}$, the overtaking time $\tau_{\rm o}$ and the nucleation time $\tau_{\rm n}$  scale proportional to $k_r^{-1}$ for increasing rescaled binding rate $k_r$. For large values in $k_r$, diffusion in bulk and surface becomes rate-limiting, leading to a plateau of all three quantities.  
\textbf{d)}
The contact angle $\theta$ has an initial decrease prior to the nucleation of a surface droplet. After its nucleation, there is significantly accelerated relaxation toward the partially wetted equilibrium state. There is a little hump in the time trace of $\theta$, which arises from the spatial-temporal change of the surface the bulk droplet is wetting.  
Parameters: $t_0= \nu^{2/3}/(\Lambda_0 k_\text{B}T)$, more see Table~\ref{tab:model_value}.
}
\label{fig:figure3b}
\end{figure*}

The area of the bulk droplet $A$ and surface droplet $A_s$ with time shift 
enable to quantitatively extract the overtaking time $\tau_{\rm o}$ of both interfaces, in addition to the nucleation and growth time, $\tau_{\rm n}$ and $\tau_{\rm g}$ (Fig.~\ref{fig:figure3b}a).
The overtaking time $\tau_{\rm o}$ corresponds to the intersection of $A$ and $A_s$.  
For $t<\tau_{\rm n}$, the bulk droplet spreads essentially on the dilute phase, while for $\tau_{\rm n}<t<\tau_{\rm o}$, the bulk droplet spreads on a composition of dense and dilute phase on the surface. 
Finally, for $\tau_{\rm o}<t<\tau_{\rm g}$, 
the surface spreads inside the surface while the bulk droplet increases its contact a bit followed by a decrease to its equilibrium surface area.

Increasing the rescaled binding rate $k_r$ shifts the contact area  
to shorter time scales 
(Fig.~\ref{fig:figure3b}b).
All three time scales, the nucleation time $\tau_{\rm n}$, the growth time $\tau_{\rm g}$ and the overtaking time $\tau_{\rm o}$ decrease algebraically proportional to $\kappa_r^{-1}$ (Fig.~\ref{fig:figure3b}c).
For large values of $\kappa_r$, the three time scales are constant. Similar to the case of complete wetting, the reason is that diffusion in surface and bulk is rate limiting.  

The contact angle $\theta$ with time shows a non-monotonous decrease toward the equilibrium value (Fig.~\ref{fig:figure3b}d). 
While the contact angle decreases most times, it increases right after nucleation of the surface droplets. This effect stems from 
a transient depletion of $\phi_s$ underneath the bulk droplet interface. Depletion arises due to the diffusion of bound molecules toward the nucleated and growing surface droplet. 
According to local equilibrium at the triple line (Eq.~\eqref{eq:young_dupre}) and following the discussion 
at the beginning of Sect.~\ref{sect:spreading_toward_complete}, a local decrease in $\phi_s$ underneath the bulk droplet interface increases the contact angle $\theta$. In other words, this transient depletion due to the nucleation of the surface droplet effectively makes the surface a bit less attractive.

\subsubsection{Scaling of the growth time $\tau_{\rm g}$ for partial wetting}

To derive the scaling behavior for the growth time under complete wetting conditions, we start by noting the arguments leading to the time evolution $d\bar{\phi}_{\rm s}/dt$ in Eq.~\eqref{eq:number_of_particles_1} can be applied here as well (Sect.~\ref{sect:taugcomplete}). 
However, the relation between $d\bar{\phi}_{\rm s}/dt$ and the binding flux need to be adjusted. 
Specifically, the position of the dense-dilute interface in the surface $X_0$ does not coincide with the droplet interface. We therefore introduce $X_{\rm d}$ to indicate the droplet interface position on the surface, where $X_{\rm d}$ itself is time dependent. The change of $\bar{\phi}_{\rm s}$ over time thus reads
\begin{equation}
    \frac{d\bar{\phi}_{s}}{dt} = -2k_{\rm r}\frac{\mu_{s}^\text{II}}{k_\text{B}T}\left(1-\frac{X_{\rm d}(t)}{X_{\rm max}}\right)\, .
\end{equation}
For fast binding $k_{r}\gg 1$, the time evolution of $X_{\rm d}(t)$ is limited by diffusion in bulk and can be considered to be slow compared to the spreading dynamics in the surface. These considerations lead us to the scaling $d X_0/dt \sim k_{r}$ and $\tau_{\rm g}\sim k_{r}^{-1}$ for the growth period.

\section{Conclusions}\label{sect:conclusion_outlook}

Our research sheds light on a relatively unexplored facet of droplet spreading in the presence of surface binding. Using irreversible thermodynamics, we obtain the continuum equations for this wetting process and study the spreading of a droplet on a surface on which the droplet components can bind. A key finding is that binding controls the spreading dynamics toward a partially and a completely wetted thermodynamic equilibrium state. 
In particular, the spreading time scales with $k_r^{-1}$, where $k_r$ is the binding rate to the surface. 
Spreading occurring on the characteristic time $k_r^{-1}$ is a result of our consideration of small systems, i.e., a droplet spreading on a finite surface. 
Preliminary studies indicate that the spreading on larger surfaces is consistent with the expected power law behaviors~\cite{Gennes1984, gennes2004capillarity}. 
However, the verification of such slowly decaying power laws is difficult to investigate when numerically solving continuum equations as this requires system sizes and simulation times beyond currently available computational resources. 
Thus, we focus on the effects of surface binding on the spreading dynamics in smaller systems.
A striking observation of our study is that binding creates a layer of droplet components on which spreading can be significantly accelerated. 
Acceleration is more pronounced if molecules remain attractive to each other after binding to the surface. This case leads to positive feedback on the wetting dynamics, which is more pronounced when more molecules are bound.  

Our work describes a mechanism that is capable of controlling the wetting dynamics via the binding of molecules to surfaces. 
By manipulating the binding rates, we demonstrate control over the nucleation and growth rates of a droplet on the surface, which gives rise to accelerated wetting dynamics.
There are other mechanisms that control wetting through changes of the surface properties, for example, adaptive wetting and reactive wetting.

{
We present a study on the wetting dynamics of a liquid droplet on a solid substrate, focusing on how the molecule-binding at the substrate surface increases the wettability of the substrate, thereby providing positive feedback for the wetting process. This situation is typical for systems undergoing  reactive wetting~\cite{Kumar_2007,Eustathopoulos2006,PRE_Zheng_1998,Hondros_1998,PRE_Sumino_2005, PRL_Sumino_2005, John_2005} and adaptive wetting~\cite{Butt2018, Hartmann2024, Hartmann2023}.
Reactive and adaptive wetting have been studied in systems where the substrate becomes either more or less wettable, or experiences alterations such as swelling, shrinking, or changes in surface chemistry.
Our work studies a molecular mechanism of  substrate properties effectively changing via molecular binding. 
This mechanism is fundamental in biological systems where proteins specifically bind receptors, altering the surface interactions with other molecules~\cite{pombogarcía2024,moser2009,harrington2021}.
Such surface modifications are also relevant in non-biological systems, as studied in metallurgy, for example~\cite{orejon2024,chobaomsup2020}.
Thus, our work may be a step towards a more unified description of complex wetting  connecting the fields of reactive and adaptive wetting with biophysical systems composed of surface-binding biomolecules. 
}

Wetting controlled via binding could be relevant for biomolecular condensates wetting membrane-bound organelles in living cells. 
For example, by enhancing the binding affinity of phase-separating proteins (phosphorylation, etc.), 
the wetting propensity of biomolecular condensates and their spreading speed 
can be accelerated by nucleating a condensate of bound proteins on the organelle surface.  
An exciting layer of complexity emerges because intracellular membranes can vary significantly in their curvature. While a membrane interaction with a small droplet can always be approximated as a flat surface, we expect membrane curvature to become more relevant with increasing droplet size.
Strikingly, condensates are expected to wet with a smaller contact angle or even completely wet the organelles if the condensate components can bind specifically to that organelle. 
Such a binding-mediated control mechanism could be crucial in regulating the communication between biomolecular condensates and membrane-bound organelles through specific feedback loops.

Future theoretical investigations should be concerned with the role of hydrodynamics during the spreading dynamics with surface binding~\cite{Anderson1998, YUE_ZHOU_FENG_2010}.
An interesting related question is whether the binding layer is an example of the thin surface film preceding the wetting dynamics, often called precursor film~\cite{hardy1919boundary, Gennes1984, leger1988precursor, xu2004molecular, popescu2012precursor}. 
Future research will delve into pattern formation on surfaces in bulk-surface systems driven by fuel-driven binding cycles~\cite{Weber_IOP_2019,bartolucci2021controlling}.

\appendix

\section{Derivation of Binding Flux}\label{APP:binding_flux}

In this section, we derive the relationship (Eq.~\eqref{eq:bc_flux_binding}) between the surface diffusion flux $\vect{j}_s$ and the binding flux $r$.
Taking the time derivative of the conserved particle number $N$ (Eq.~\eqref{eq:tot_N}, and $dN/dt=0$), we get
\begin{equation}
    0= \int_S dS \left(\frac{\partial \phi_s }{\partial t} \right)/\nu_s + \int_V dV \left( \frac{\partial\phi}{\partial t} \right) /\nu \, . 
\end{equation}
Using the conservation laws (Eqs.~\eqref{eq:cons_laws_full_sodel}) and the divergence theorem, 
\begin{equation}
0= \int_S dS \left( r/\nu_s - \vect{n} \cdot \vect{j} /\nu \right)  - \int_{\partial V} dS \, \vect{n} \cdot \vect{j}/\nu  - \int_{\partial S} dl \, \vect{t} \cdot \vect{j}_s/\nu_s  \, .
\end{equation}
This condition is fulfilled when Eq.~\eqref{eq:bc_flux_binding} is satisfied, together with 
Eq.~\eqref{eq:bc_flux_nobinding}. For $\partial S$, we consider periodic boundary conditions. 

\section{Initialisation of wetting studies and equilibrium contact angle for partial and complete wetting}\label{APP:phase_diag}

\begin{figure*}[bt]
\centering
{\includegraphics[height=5.5cm]{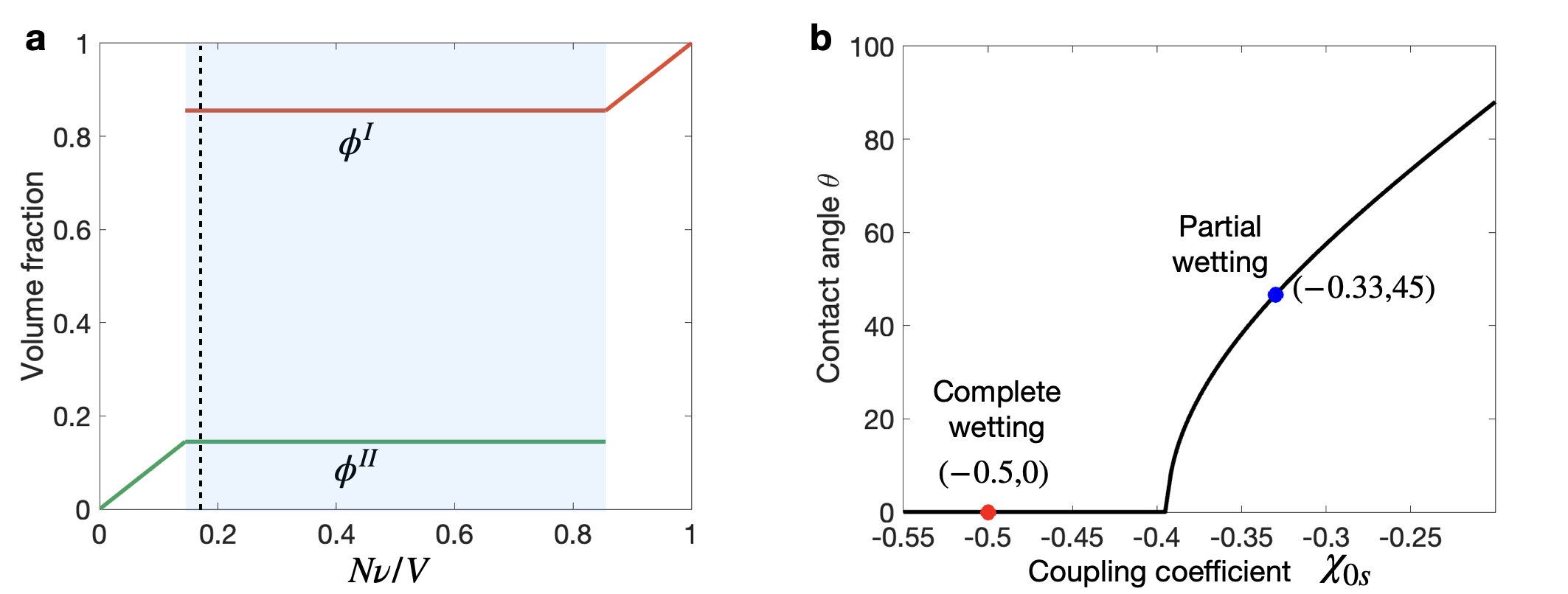}}
\caption{
\textbf{{Parameter initialization in the numerical simulations.}} 
\textbf{a)} To study the wetting dynamics, a bulk droplet is initialized with a volume $V^\text{I}$ and  volume fractions in the molecule-rich  (I) and molecule-poor (II) phase corresponding to the phase diagram without binding ($\phi_s=0$).
For all studies, the conserved total volume fraction is $N\nu/|V|={0.17}$, which is indicated by the vertical dashed line.
\textbf{b)} The wetting dynamics approaches thermodynamic equilibrium with an equilibrium contact angle $\theta$. 
This value is determined by the interaction parameter between bulk and surface, $\chi_{0s}$.
We consider two cases: relaxation toward a completely wetted state ($\chi_{0s}=-0.5$) and a partially wetted state ($\chi_{0s}=-0.33$).
We note that we calculate the equilibrium values for $\phi_s$, $\phi$ with respect to difference interaction parameter strength $\chi_{0s}$ using the equilibrium conditions in an ensemble where the chemical potential $\mu$ is fixed (see details in~\cite{NJPpaper2021}). The contact angle $\theta$ is obtained by the law of Young–Dupré at equilibrium.
}
\label{fig:appendix_phase_diagram1}
\end{figure*}

For all wetting studies, a single bulk droplet is initialized right on top of the surface where no molecules are initially bound ($\phi_s(\vect{x}_\parallel, t=0)=0$). 
The volume fractions inside (I) (molecule-rich phase) and outside (II) (molecule-poor phase) are chosen to be homogeneous at $t=0$ and in accordance with the equilibrium volume fractions in the absence of surface binding, $\phi^\text{I}$ and $\phi^\text{II}$ (Fig.~\ref{fig:appendix_phase_diagram1}a). 
Moreover, at $t=0$, we also use the equilibrium value for bulk droplet volume, $V^\text{I}(t=0)=(N\nu -\phi^\text{II})/(\phi^\text{I}-\phi^\text{II})$. Note that for all studies, the conserved total volume fraction is $N\nu/|V|={0.17}$. The position of the droplet center is chosen such that wetting dynamics get initiated by the bridge formation of a molecule-rich phase in the absence of fluctuations.

At large times, the wetting dynamics reaches the corresponding thermodynamic equilibrium state. In our work, we studied the dynamics towards to completely wetted state (Sect.~\ref{sect:spreading_toward_complete}) and a partially wetted state (Sect.~\ref{sect:spreading_toward_partial}). For both cases, the used parameters are given in Table~\ref{tab:model_value}, except the interaction parameter between bulk and surface, $\chi_{0s}$. 
The more negative this parameter, the stronger the attraction between bound molecules and the bulk molecules adjacent to the surface.
In other words, decreasing $\chi_{0s}$ towards more negative values leads to a transition between a partially wetted to a completely wetted equilibrium state (Fig.~\ref{fig:appendix_phase_diagram1}b).
For our studies, we employ $\chi_{0s}=-0.5$ leading to a completely wetted state with a contact angle $\theta=0$ and a homogeneous surface area fraction $\phi_s=0.9717$ at thermodynamic equilibrium. 
For $\chi_{0s}=-0.33$, the bulk droplet partially wets the surface. 
The corresponding equilibrium contact angle $\theta\simeq\pi/4$ and the surface  area fraction is slightly different underneath the molecule-rich (I) ($\phi_s=0.9156$) and molecule-poor bulk phase (II) ($\phi_s=0.8769$).
{We note that we calculate the equilibrium values for $\phi_s$ with respect to difference interaction parameter strength $\chi_{0s}$ using the equilibrium conditions in an ensemble where the chemical potential $\mu$ is fixed (see details in~\cite{NJPpaper2021}). The contact angle $\theta$ is obtained by the law of Young–Dupré.}

\section{Nucleation of surface droplet by bulk droplet}\label{APP:nucleation_surface_drop}

\begin{figure*}[bt]
 \centering
 {\includegraphics[height=6cm]{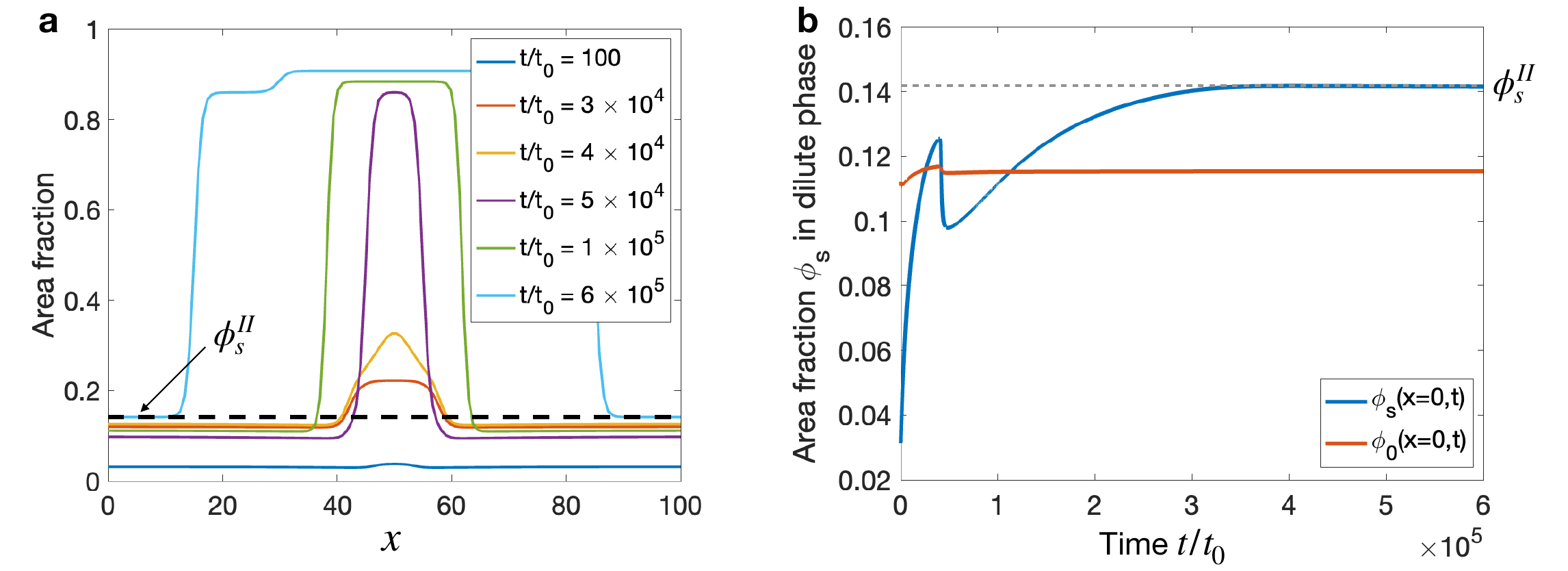}}
 \caption{
 \textbf{{Heterogeneous nucleation of a surface droplet induced by a bulk droplet situated on top of the surface}.} 
 \textbf{a)} The area fraction $\phi_s$ of surface-bound molecules increases and peaks below the center of the bulk droplet {due to the heterogeneous surface binding}. Once this peak area fraction {exceeds} the equilibrium value $\phi^\text{II}_s$ (black dash line, sketch see Fig.~\ref{fig:schematic}c), a droplet at the surface center is nucleated. 
 \textbf{b)} { The surface-bound area fraction $\phi_s$ (blue solid line) and the bulk volume fraction at the bottom boundary $\phi|_{z=0}$ in the dilute phase change non-linearly. Initially, the nucleation of a surface droplet at the center leads to a decrease in the area fraction value far from the center. On longer times, the area fraction $\phi_s$ far from the surface droplet at $x = 0$ approaches the equilibrium area fraction $\phi_s^\text{II}$ (dashed line in plot (a)) from below. The bulk volume fraction at the surface $\phi|_{z=0}(x=0, t)$ shows a similar as the area fraction of the bound molecules $\phi_s(x=0, t)$ but with less magnitude variation due to the attractive coupling between $\phi_s(x=0, t)$ and $\phi|_{z=0}(x=0, t)$.
We note that when the interface of the surface droplet reaches the system size, there is a sudden increase in the area fraction to the thermodynamic equilibrium value (dense phase, not shown).
}
 For both figures,  $\chi_{0s} = -0.33$. Further parameter values are listed in Table~\ref{tab:model_value}. }
 \label{fig:phi_s_t}
 \end{figure*}

After bridge formation on the slightly repulsive surface ($\omega_0>0$), the binding of molecules from the bulk to the surface enhances the wetting propensity of the bulk droplet. Binding makes the surface effectively more attractive through the  interactions between bound molecules and bulk molecules adjacent to the surface ($\chi_{0s}<0$).
Interestingly, the increase of bound molecules is more pronounced underneath the center of the bulk droplet (Fig.~\ref{fig:phi_s_t}a). 
Once the local area fraction $\phi_s$ exceeds the equilibrium area fraction $\phi_s^\text{II}=0.1415$ (sketch see Fig.~\ref{fig:schematic}c), a surface droplet gets nucleated. Note that only the position centered underneath the bulk droplet crosses the equilibrium value $\phi_s^\text{II}$, while the surface domain far away and closer to the boundaries remains undersaturated. 
The corresponding area fractions far away from the surface droplet approach $\phi_s^\text{II}$ from below (Fig.~\ref{fig:phi_s_t}b). This study shows that the bulk droplet indeed nucleates the formation of a droplet on the surface.

\section{Parameters used for wetting studies}\label{APP:parameters}

In our wetting studies, we consider two values of the interaction parameter between bulk and surface $\chi_{0s}=\{-0.5,-0.33\}$, corresponding to the cases where the system approaches complete or partial wetting at thermodynamic equilibrium. We also vary the dimensionless binding rate $k_r$ (definition see Eq.~\eqref{eq:nondim_parameters}).
The remaining parameters are kept fixed for all presented studies.
Such parameters are summarized in Table~\ref{tab:model_value}.

\section{Numerical study in three-dimensional cylindrical coordinates with axial symmetry}\label{APP:3D_cylinder}
\begin{figure*}[h!]
\centering
{\includegraphics[height=12cm]{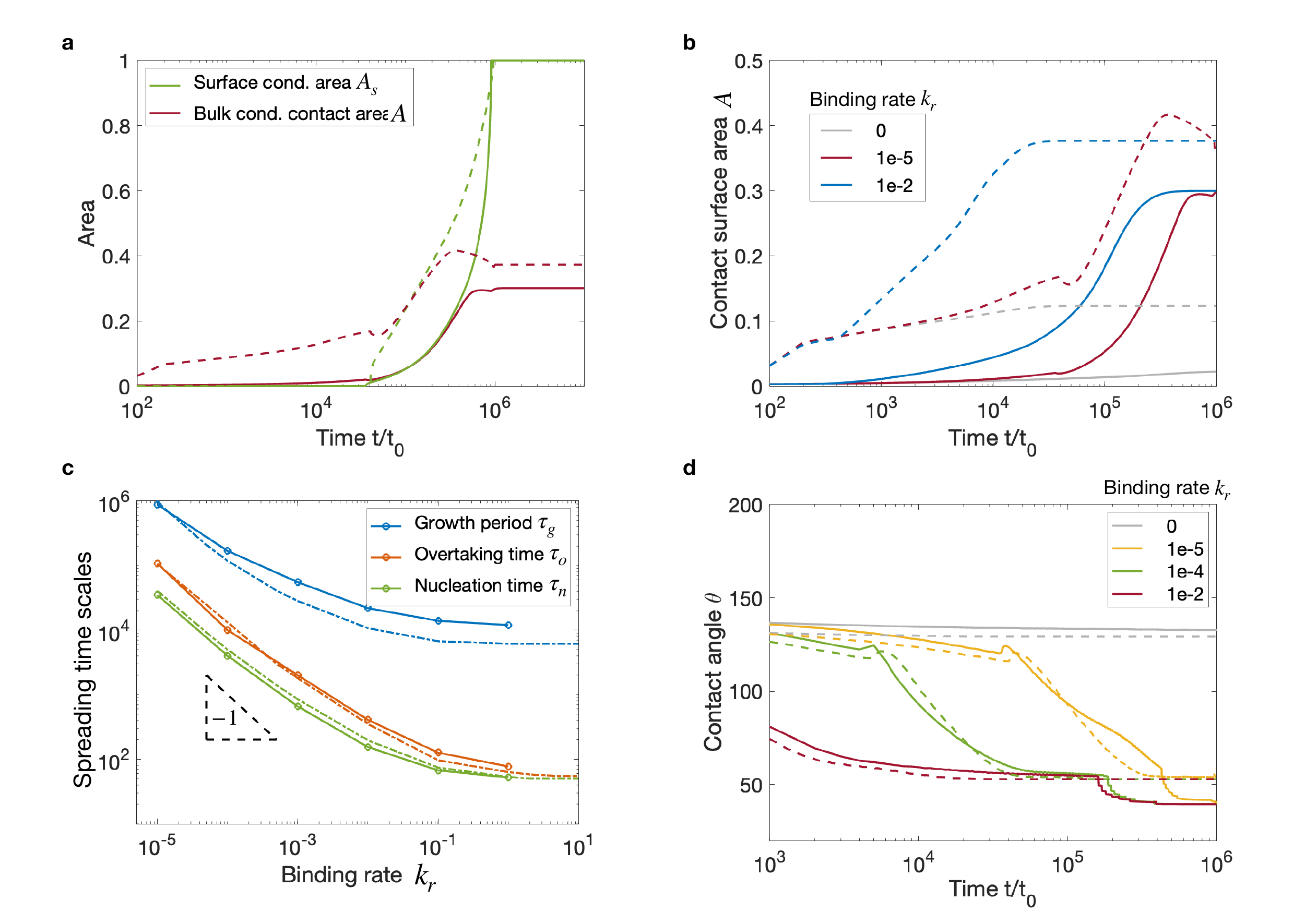}}
\caption{
\textbf{Spreading dynamics towards a partially wetted state in three-dimensional cylindrical coordinates with rotational symmetry (axial symmetric case).} In all the plots presented here, solid lines denote the results in three-dimensional (3D) cylindrical coordinates and dashed lines the referenced results in two-dimensional (2D) Cartesian coordinates (see Fig.~\ref{fig:figure3b}).
\textbf{a)} and \textbf{b)} compare the contact area of the bulk droplet and the area of the surface droplet between 2D and 3D, illustrating how dimensionality affects the spreading behavior. 
\textbf{c)} 2D and 3D studies give almost the same spreading time scales ($\tau_g, \tau_o, \tau_n$). \textbf{d)} We find a similar dynamic behavior between 2D and 3D but different equilibrium profiles in contact angle $\theta$. 
}
\label{fig:figure3b_comparison}
\end{figure*}

To explore whether the qualitative results for the planar setting are still valid in three dimensions (3D), we consider a rotational symmetric problem in cylindrical coordinates. The Laplace operator in cylindrical coordinates for a axial symmetric system reads:
\begin{align}
    \Delta = d^2/dr^2 + (1/r) d/dr + d^2/dz^2.
\end{align}
{We repeat the numerical calculations with the parameter values in Section~\ref{subsec:partial_wetting}, and compare the results with the ones in two dimensional (2D) Cartesian coordinates~(See Fig.~\ref{fig:figure3b_comparison} for details).} {We found that 2D and 3D studies share almost same spreading time scales ($\tau_g, \tau_o, \tau_n$; see Fig.~\ref{fig:figure3b_comparison}c) and similar dynamic behavior in contact angle $\theta$, but with notable differences in the equilibrium profiles (Fig.~\ref{fig:figure3b_comparison}d), as well as in the bulk droplet contact area $A$ at equilibrium (Fig.~\ref{fig:figure3b_comparison}(a,b)). Still, these results indicate that qualitative results for the spreading dynamics with surface binding are robust between two and three-dimensional domains.}

\begin{center}
\begin{table}
    \begin{tabular}{| l | c | c | }
    \hline
    \textbf{Parameter name} & \textbf{Symbol} & \textbf{rescaled value } 
    \\ \hline
     interaction coefficient at the surface & $\chi_{s}$   & 2.5 \\
    \hline
     interaction coefficient in bulk & $\chi$  & 2.5 \\ 
     \hline
    binding energy per unit area& $\omega_0$ & 0.17 \\
    \hline
    internal free energy coefficient of molecule in the bulk& $\omega$ & 2.5 \\
    \hline
    internal free energy coefficient of molecules at the surface& $\omega_s$ & 2.5 \\
    \hline
    interaction coefficient accumulating at the surface& $\chi_{00}$ & 0 \\
    \hline
    diffusion coefficient in the membrane & $D_{\rm s}$ & 1\\
     \hline
      gradient coefficient of molecule in the membrane & $\kappa_s$ &   1 \\
     \hline
      gradient coefficient of molecule in the bulk & $\kappa$ & 1 \\
     \hline
      domain size of the bulk & $L_x \times L_y$  & 100$\times$30 \\
     \hline
    \end{tabular}
    \caption{\label{tab:model_value}\textbf{Model parameters and their dimensionless values considered for the studies on wetting dynamics.}}
    \end{table}
\end{center}

\section*{Conflicts of interest}
There are no conflicts to declare.

\section*{Acknowledgements}
{We thank all the anonymous reviewers for very constructive feedback on the manuscript.} We thank S.\ Gomez for carefully reading our manuscript and giving feedback, and G.\ Granatelli for improving the illustration (Fig.~\ref{fig:sketch}).
We thank G.\ Bartolucci, J.\ Bauermann, L.\ Hubatsch, S.\ Bo, S.\ Laha, T.\ Harmon, I.\ LuValle-Burke, and D.\ Sun for insightful discussions.
We thank David Zwicker for his valuable comments on calculating contact angles.
X.\ Zhao acknowledges the ``FoSE New Researchers Grant'' of the University of Nottingham Ningbo China for financial support.
A.\ Honigmann and C.\ Weber acknowledge the SPP 2191 ``Molecular Mechanisms of Functional Phase Separation'' of the German Science Foundation for financial support and for providing an excellent collaborative environment. 
\bibliography{ms} 

\end{document}